\documentclass[a4paper,11pt]{article}
\pdfoutput=1 
\usepackage{soul}

\usepackage{jcappub} 

\usepackage[T1]{fontenc} 
\usepackage[normalem]{ulem}

\newcommand{\pos}{_\mathbf{r}}
\newcommand{\ppos}{_{\mathbf{r}+\mathbf{r'}}}
\newcommand{\na}{\bar{n}_a}
\newcommand{\nb}{\bar{n}_b}
\newcommand{\nc}{\bar{n}_c}
\newcommand{\nd}{\bar{n}_d}

\newcommand{\ab}{_{ab}}
\newcommand{\abc}{_{abc}}
\newcommand{\abd}{_{abd}}
\newcommand{\acd}{_{acd}}

\newcommand{\abcd}{_{abcd}}
\newcommand{\bc}{_{bc}}
\newcommand{\bd}{_{bd}}
\newcommand{\ac}{_{ac}}
\newcommand{\cd}{_{cd}}
\newcommand{\ad}{_{ad}}

\newcommand{\ra}{_{\mathbf{r}_a}}
\newcommand{\rb}{_{\mathbf{r}_b}}
\newcommand{\rc}{_{\mathbf{r}_c}}
\newcommand{\rd}{_{\mathbf{r}_d}}
\newcommand{\rab}{_{\mathbf{r}_{ab}}}
\newcommand{\rabc}{_{\mathbf{r}_{abc}}}
\newcommand{\rabd}{_{\mathbf{r}_{abd}}}
\newcommand{\racd}{_{\mathbf{r}_{acd}}}
\newcommand{\rbcd}{_{\mathbf{r}_{bcd}}}
\newcommand{\rabcd}{_{\mathbf{r}_{abcd}}}
\newcommand{\rbc}{_{\mathbf{r}_{bc}}}
\newcommand{\rbd}{_{\mathbf{r}_{bd}}}
\newcommand{\rac}{_{\mathbf{r}_{ac}}}
\newcommand{\rcd}{_{\mathbf{r}_{cd}}}
\newcommand{\rad}{_{\mathbf{r}_{ad}}}
\newcommand{\mom}{_\mathbf{k}}

\newcommand{\mrm}{\mathrm}
\newcommand{\mbf}{\mathbf}
\newcommand{\gal}{^\mrm{g}}
\newcommand{\rnd}{^\mrm{s}}
\newcommand{\an}{\bar{n}}
\newcommand{\DD}{\delta^\mrm{D}}
\newcommand{\mr}[1]{\mathrm{#1}}

\usepackage{cancel}
\usepackage{multirow}
\usepackage{mathtools}

\title{\boldmath 
Integrated trispectrum detection from BOSS DR12 NGC CMASS}

\author[a,b]{Davide Gualdi,}
\author[a,c]{and Licia Verde}

\affiliation[a]{Institut de Ci\`encies del Cosmos, University of Barcelona, ICCUB, Barcelona 08028, Spain}
\affiliation[b]{Institute of Space Studies of Catalonia (IEEC), E-08034 Barcelona, Spain}
\affiliation[c]{Instituci\'o Catalana de Recerca i Estudis Avan\c{c}ats, Passeig Llu\'is Companys 23, Barcelona 08010, Spain}

\emailAdd{dgualdi@icc.ub.edu}
\emailAdd{liciaverde@icc.ub.edu}

\abstract{ We present the first detection of the integrated trispectrum ({\it i-trispectrum}) monopole and quadrupoles signal from BOSS CMASS NGC DR12. Extending the FKP estimators formalism to the Fourier transform of the four-point correlation function, we test shot-noise subtraction, Gaussianity of the i-trispectrum data-vector, significance of the detection and similarity between the signal from the data and from the galaxy mock catalogues used to numerically estimate the covariance matrix. Using scales corresponding to modes from minimum $k_\mathrm{min}=0.03\,h/\mathrm{Mpc}$ to maximum $k_\mathrm{max}=0.15\,h/\mathrm{Mpc}$,   we find a detection in terms of   distance from the null hypothesis  of $(10.4,5.2,8.3,1.1,3.1)$ $\sigma$-intervals for the i-trispectrum monopole $\mathcal{T}^{(0)}$ and quadrupoles $(\mathcal{T}^{(2000)},\mathcal{T}^{(0200)},\mathcal{T}^{(0020)},\mathcal{T}^{(0002)})$ respectively. This quantifies the presence of the physical signal of the  four-points statistics on BOSS data. For completeness
the same analysis is also performed for power spectrum and bispectrum, both monopoles and quadrupoles.
}

\begin{document}
\maketitle
\flushbottom

\section{Introduction}
The power spectrum (the two-point correlation function) is and has long been the workhorse summary statistics used to interpret clustering in  galaxy surveys. For Gaussian initial conditions it captures  the bulk of the cosmologically relevant information content. Due to non-linearity induced by gravitational collapse, however,  signal leaks from 2-point (2pt) statistics into higher-order moments of the density field e.g., \citep{Peebles1980}.
This is the reason why higher-order statistics of the galaxy density field are one of the most promising routes to extract additional cosmological information beyond the power spectrum from current and future large scale structure (LSS) surveys (e.g., DESI\footnote{\url{http://desi.lbl.gov}} \citep{Levi:2013gra}; Euclid \footnote{\url{http://sci.esa.int/euclid/}} \citep{Laureijs:2011gra}; PFS \footnote{\url{http://pfs.ipmu.jp}} \citep{Ellis:2012rn}; SKA\footnote{\url{https://www.skatelescope.org}} \citep{Bacon:2018dui}; LSST\footnote{\url{https://www.lsst.org/}} \citep{Abell:2009aa} and WFIRST\footnote{\url{https://www.cosmos.esa.int/web/wfirst}} \cite{Green:2012mj}). 

The 3pt correlation function and its Fourier transform - the bispectrum - have been widely studied in the literature as the lowest higher-order statistics able to lift parameters degeneracies present at the 2pt level \citep{Groth:1977gj,1975ApJ...196....1P,1982ApJ...259..474F,Fry:1983cj,Fry:1992vr,Fry:1993bj,Matarrese:1997sk,Verde:1998zr,Verde:2001sf,Scoccimarro:1997st,Scoccimarro:2000ee,Scoccimarro:1999ed,Scoccimarro:2000sp,Oddo:2019run,Barreira:2019icq,Agarwal:2020lov,Biagetti:2021tua,Alkhanishvili:2021pvy} and therefore to capture additional information \cite{Yankelevich:2018uaz,Samushia:2021ixs,Oddo:2021iwq}.
In particular, by employing the bispectrum in addition to the power spectrum, constraints on neutrino masses sum can be improved \cite{Ruggeri:2017dda,Coulton:2018ebd,Hahn:2019zob,Hahn:2020lou,Kamalinejad:2020izi}, together with baryonic acoustic oscillations (BAO) \cite{Pearson:2017wtw,Child:2018klv,Slepian:2016kfz} and relativistic effects \cite{GilMarin:2011xq,Bartolo:2013ws,Bellini:2015wfa,Bertacca:2017dzm,DiDio:2018unb}. 

The state-of-the-art on measurement and interpretation of the bispectrum from galaxy surveys  data is represented by the cosmological analyses performed on BOSS SDSS III survey data for both isotropic bispectrum \cite{Gil-Marin:2014sta,Gil-Marin:2016wya,Philcox:2021kcw} and 3pt correlation function \cite{Slepian:2015hca}. Methods to measure and model the anisotropic bispectrum signal have also been introduced  \cite{Scoccimarro:2015bla} and applied to data \cite{Sugiyama:2018yzo}, with studies on the relative additional signal \cite{Sugiyama:2019ike} and forecasted reduction in parameter constraints \cite{Gagrani:2016rfy,Gualdi:2020ymf}.

Cosmic microwave background (CMB) experiments and analyses have exploited the signal from the 4pt correlation function Fourier transform, the trispectrum, to constrain primordial non-Gaussianity 
\cite{kunzetal2001,Komatsu:2002db,deTroia2003,Munshi:2009wy,Kamionkowski:2010me,Izumi:2011di,Regan_2015, Feng:2015pva,Fergusson:2010gn,Smith:2015uia,Namikawa:2017uke,PlanckTrispectrum18, Akrami:2019izv}.
CMB analyses have demonstrated  the trispectrum constraining power for primordial non-Gaussianities, with implications for confirming or ruling out single/multi field inflation models \cite{Bartolo:2004if}. Since the late-time 3D matter field trispectrum, by definition, contains more modes than the primordial 2D CMB counterpart, as pointed out by \cite{Verde:2001pf}, the LSS trispectrum can be a powerful tool in deriving late-time constraints on primordial non-Gaussianity.

However, in the  late-time Universe, the challenge of measuring and modelling the same statistic for a three-dimensional field instead of a two-dimensional map is much harder and this has contributed to reduce the interest in the LSS trispectrum \cite{Verde:2001pf,Cooray:2008eb,Lazeyras:2017hxw,Bellomo:2018lew}, with few measurements from data and simulations \cite{Fry1978,Suto:1993ua,Sabiu:2019kbh}. Recently, the 4pt correlation function in configuration space was detected from BOSS data \cite{Philcox:2021hbm}.

In terms of modelling, the effective field theory of LSS formalism was applied to the trispectrum by \cite{Bertolini:2016bmt} and recently calibrated at 1-loop in \cite{Steele:2021lnz}. An angular coordinates formalism has been recently introduced in \cite{Lee:2020ebj} and a compression for weak-lensing in \cite{Munshi:2021uwn}, while in the presence of a primordial trispectrum, the correction to the non-Gaussian linear bias was derived by \cite{Lazeyras:2015giz}.

An estimator for an integrated version of the 3D LSS trispectrum was proposed first by \cite{Sefusatti:2004xz} and in \cite{Gualdi:2020eag} we measured its isotropic signal from the Quijote simulations suite \cite{Villaescusa-Navarro:2019bje}. Ref.~\cite{Gualdi:2020eag} also  computed the corresponding analytical theoretical model for this summary statistic, proving the potential it holds for constraining primordial non-Gaussianity. This was further confirmed by performing a more realistic joint-analysis in \cite{Gualdi:2021yvq} including both monopole and quadrupoles of power spectrum, bispectrum and i-trispectrum measurements from simulations on a much larger parameter set comprising nuisance and cosmological parameters normally considered in realistic  LSS clustering analyses.

This work presents the next step toward the goal of  employing 4pt statistics in the analysis of real data: the i-trispectrum monopole and quadrupoles detection from BOSS SDSS CMASS NGC DR12 galaxy field \cite{BOSS:2016wmc}. To achieve this, we extend the FKP estimator formalism \cite{Feldman:1993ky} to the 4pt-level statistics and   implement it in our measuring pipeline.
We measure the statistics (power spectrum, bispectrum and i-trispectrum monopoles and quadrupoles) from BOSS data and 2048 realisations of the Patchy Mocks \cite{Kitaura:2015uqa,Rodriguez-Torres:2015vqa}. First, we test the shot-noise subtraction and the (approximate) Gaussianity of the i-trispectrum data-vector. We then use the signal-to-noise ratio to quantify the presence and strength of the signal in the  survey data. At the same time we test the similarity between the i-trispectrum measured from data and from mock catalogues.

The paper is organised as follow: in Section \ref{sec:data_and_mocks} the description of the data and galaxy mock catalogues used in the analysis is reported. Section \ref{sec:fkp_estimators} summarises the FKP formalism and the extension to the i-trispectrum together with testing the data-vector's Gaussianity. The quantities used to evaluate the signal detection are listed in Section \ref{sec:signal_detection}. The results first include a shot-noise subtraction test in Section \ref{sec:shot_noise_test} and subsequently report the detection in Section \ref{sec:detection_results} together with an overall study including power spectrum and bispectrum looking at the signal-to-noise ratio as a function of scale. We conclude in Section \ref{sec:conclusions}.
After deriving the FKP estimator at 4pt level in Appendix \ref{sec:fkp_estimators_appendix}, in Appendix \ref{sec:p_b_results} the equivalent results for power spectrum and bispectrum are reported. In Appendix \ref{sec:remove_quads} the criteria to remove quadrilaterals configurations whose i-trispectrum signal is dominated by the convolution with the survey window function is presented.

\section{Methodology}
\label{sec:methodology}

\subsection{Data and Mocks}
\label{sec:data_and_mocks}
We use the CMASS NGC galaxy sample (with redshift cuts such that $0.43\leq z \leq 0.70$) of the Baryon Oscillation Spectroscopic Survey (BOSS \cite{BOSS:2012dmf}) which is part of the Sloan Digital Sky Survey III \cite{SDSS:2011jap}. From the final DR12 data-release \cite{BOSS:2016wmc}, for the chosen redshift interval we use 605522 galaxies with an effective redshift $z_\mrm{eff}=0.5447$. To correct observational known systematic errors we implement the standard weighting prescription
\begin{eqnarray}
w = w_\mrm{sys}\,\left(w_\mrm{rf}+w_\mrm{fc}-1\right)\,,
\end{eqnarray}
\noindent where $w_\mrm{sys}$ accounts for target density variations, $w_\mrm{rf}$ for redshift failures and $w_\mrm{fc}$ for fiber collision.

To accurately estimate the covariance matrix, a numerical approach is needed and for this it is necessary to employ a large suite of mock galaxy catalogues. These are different mock realizations of the same region of the Universe, which, to reduce computational cost, instead of being based on standard N-body simulations, are  based on approximate methods such as second-order Lagrangian perturbation theory \cite{Scoccimarro:2001cj,Manera:2012sc} or augmented Lagrangian perturbation theory as described in \cite{Kitaura:2012tj}. Such approximations have been shown to be well suited to correctly model the field in the linear and quasi-linear regimes and to provide sufficiently accurate estimates of covariance matrices. However one should bear in mind that  the approximations involved may not be sufficient to  accurately model the signal in detail \cite{Chuang:2014vfa} especially when pushing closer to the non-linear regime.

In this work we use 2048 realisations of the MultiDark Patchy BOSS DR12 mocks by \cite{Kitaura:2015uqa,Rodriguez-Torres:2015vqa} which have the underlying cosmology: $\Omega_{\Lambda} = 0.693$, $\Omega_{\mrm{m}}(z=0) = 0.307$, $\Omega_{\mrm{b}}(z=0) = 0.048$, $\sigma_8(z=0)=0.829$, $n_{\mrm{s}} = 0.96$, $h = 0.678$. Being very close to \textit{Planck15} analysis results \cite{Planck:2015fie}, we use the same parameters values as fiducial cosmology to convert redshifts into distances also for BOSS DR12 CMASS NGC data.
These mocks include survey selection effects, light-cone evolution as a function of redshift, and galaxy biasing properties that result in a close match with the power spectrum measured from BOSS DR11$\&$DR12 galaxy samples up to $k\sim0.3 \,h/\mrm{Mpc}$. These mocks accurately modelled the covariance matrix also for 3pt statistics \cite{Gil-Marin:2016wya,Slepian:2016kfz}. In this work we show that Patchy mocks contain an i-trispectrum signal in very good agreement in terms of both shape and amplitude to the one measured from BOSS data. This will justify using the mock measurements to estimate the i-trispectrum covariance matrix.

For the Patchy Mocks we use the weighting scheme $w=w_\mrm{veto}\times w_\mrm{fc}$, where $w_\mrm{veto}$ is the weight for the veto mask. As explained in Appendix \ref{sec:fkp_estimators_appendix},
we do not employ the FKP weights $w_\mrm{FKP}$, which are used to optimise measurements from survey regions with different densities.
Being these weights usually significantly different from unity, the approximations we made to subtract the shot-noise using measured quantities would not hold for the chosen $k$-range.

\begin{figure}[tbp]
\centering 
\includegraphics[width=1.\textwidth]
{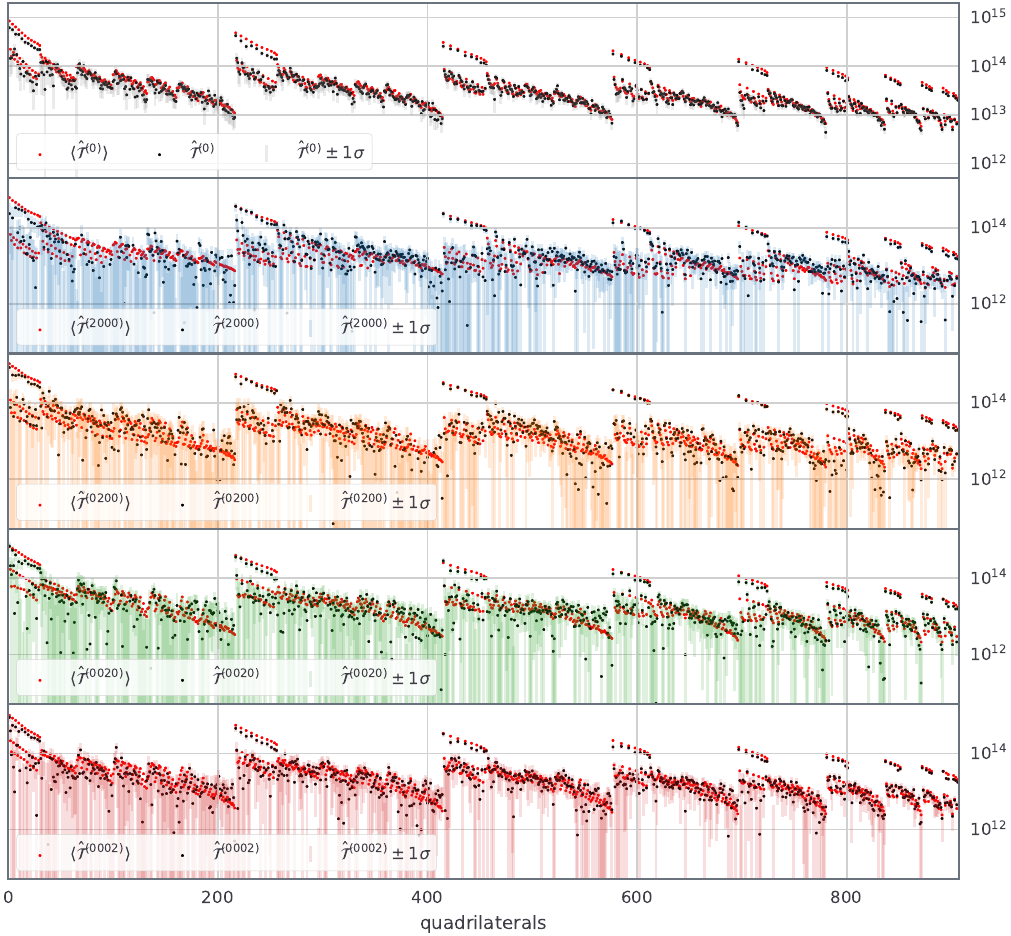}
\caption{\label{fig:T_det_all_quads}
Measurements of the i-trispectrum monopole and quadrupoles from BOSS DR12 CMASS NGC data for all the skew-quadrilaterals found for the chosen scale-cut and $k$-binning size. The error-bars are obtained from the covariance estimated using 2048 realisations of the Patchy Mocks and are centered around the black dots which are the  measurements from BOSS data. For comparison, the red dots show the average measurements of the same quantities from the Patchy Mocks. For certain quadrilaterals the i-trispectrum signal is significantly larger than for neighbouring configurations. Comparing with our results obtained in \cite{Gualdi:2020ymf,Gualdi:2021yvq} we know that this is due to the survey window effect. An a-posteriori procedure to remove these configuration is outlined in Appendix \ref{sec:remove_quads} where in Figure \ref{fig:T_det_sel_quads} we also show the resulting data-vector without the removed quadrilaterals.
}
\end{figure}

\subsection{FKP Estimators}
\label{sec:fkp_estimators}

To measure density field summary statistics it is necessary to employ estimators which help mitigate and account for 
the effects induced on the signal by the surveys window function. Each survey is indeed characterised by a specific choice for the angular mask, the redshift cut and target selection criteria. 

The workhorse estimator for the power spectrum was developed in \cite{Feldman:1993ky} (known in the literature as "FKP" estimator from the authors's surnames) and  solves these complications by introducing a synthetic catalogue with the same properties of the observed field in terms of angular mask, redshift cuts and objects density variations but without any clustering, i.e., the objects have a random distribution. This allows  one to define a proxy for the over-density field
\begin{equation}
F(\mbf{r})= w_\mrm{g}(\mbf{r})n_\mrm{g}(\mbf{r}) - \alpha\,w_\mrm{s}(\mbf{r})n_\mrm{s}(\mbf{r})\,,
\end{equation}
\noindent where $n_\mrm{g}$ and $n_\mrm{s}$ are the densities of the observed tracers and synthetic objects at the position $\mbf{r}$, with $w_\mrm{g}$ and $w_\mrm{s}$ being the associated weights (the difference between $w_\mrm{g}$ and $w_\mrm{s}$  is specific to each data-set). Usually a much larger number of synthetic objects is used to reduce to a minimum the shot-noise due to the random catalogue, with $\alpha$ being the ratio between the total number of observed tracers $N_\mrm{g}$ and random particles $N_\mrm{s}$. In our case for both data and mocks we use $\alpha=0.01$.

The anisotropic signal about the line of sight (LOS) induced by redshift space distortions (RSD) \cite{Kaiser:1987qv} can be captured in summary statistics such as the power spectrum by weighting the estimator using Legendre polynomials 
as functions of  the angle between the $k$-vector and the object's position $\mbf{r}$. The Yamamoto power spectrum estimator \cite{Yamamoto:2005dz} for example,  considers the angle between the $k$-vector and the average positions of two objects $\mbf{r} = (\mbf{r}_1+\mbf{r}_2)/2$. A much faster numerical implementation using Fast Fourier Transforms (FFTs) was introduced in \cite{Bianchi:2015oia}, with the choice of associating the LOS dependence only to one of the considered objects. This allows one to define in Fourier space the quantity  
\begin{eqnarray}
\label{eq:an_fourier}
A_n(\mbf{k})=\int d\mbf{r}^3 (\hat{\mbf{k}}\cdot\hat{\mbf{r}})^n\,F(\mbf{r})\,e^{i\mbf{k}\mbf{r}}\,,
\end{eqnarray}
\noindent which, as explained in \cite{Bianchi:2015oia}, can be used to construct estimators for different multipoles of order $n$ for the power spectrum. Similarly, Ref.~\cite{Scoccimarro:2015bla} extends this formalism to the bispectrum, which was then implemented in the BOSS analysis 
for the bispectrum monopole \cite{Gil-Marin:2016wya}. In this work we extend this formalism to the 4pt level and include each of the considered statistics' quadrupoles. Using Equation~\ref{eq:an_fourier} we define two quantities 
\begin{eqnarray}
\label{eq:f0_and_f2}
F_0(\mbf{k}) = A_0(\mbf{k})\,,\quad\quad F_2(\mbf{k}) = 3A_2(\mbf{k}) - A_0(\mbf{k})\,.
\end{eqnarray}

\noindent Then, as derived in Appendix \ref{sec:fkp_estimators_appendix}, the following estimators can be adopted for 2pt, 3pt and 4pt correlators in Fourier space to measure respectively power spectrum, bispectrum and i-trispectrum monopoles  plus shot-noise terms

\begin{align}
\label{eq:fkp_continuous_estimators}
\langle F_0(\mbf{k})F_0(\mbf{k})\rangle 
&=
\mathcal{I}_{22}^{-1}\Bigg\{
\int \dfrac{d\mbf{q}^3}{(2\pi)^3}P^{(0)}(\mbf{q})
\mathcal{W}^{11}_{\mbf{k}-\mbf{q}}
\mathcal{W}^{11,*}_{\mbf{k}-\mbf{q}}
+(1+\alpha )\mathcal{I}_{21}\Bigg\}\,,
\notag \\
\langle F_0(\mbf{k}_1)F_0(\mbf{k}_2)F_0(-\sum_{i=1}^2\mbf{k}_i)\rangle 
&=
\mathcal{I}^{-1}_{33}\Bigg\{
\int\dfrac{d\mbf{k}_i^3d\mbf{k}_j^3}{(2\pi)^6}
B^{(0)}(\mbf{k}_i,\mbf{k}_j)\,
\mathcal{W}^{11}_{\mbf{k}_1-\mbf{k}_j}
\mathcal{W}^{11}_{\mbf{k}_2-\mbf{k}_i}
\mathcal{W}^{11}_{\mbf{k}_i+\mbf{k}_j-\mbf{k}_1-\mbf{k}_2}
\notag\\
&
+
\int\dfrac{d\mbf{k}^3}{(2\pi)^3}P^{(0)}(\mbf{k})\,
\left[
\mathcal{W}^{11}_{\mbf{k}_1-\mbf{k}} 
\mathcal{W}^{21}_{\mbf{k}-\mbf{k}_1}
+2\,\mrm{cyc.}
\right]
+
(1-\alpha^2)\,\mathcal{I}_{31}\Bigg\}
\,,
\notag \\
\langle F_0(\mbf{k}_1)F_0(\mbf{k}_2)F_0(\mbf{k}_3)F_0(-\sum_{i=1}^3\mbf{k}_i)\rangle 
&=
\mathcal{I}^{-1}_{44}\Bigg\{
\int\dfrac{d\mbf{k}_i^3d\mbf{k}_j^3d\mbf{k}_l^3}{(2\pi)^9}
T^{(0)}(\mbf{k}_i,\mbf{k}_j,\mbf{k}_l)
\notag \\
&\times
\mathcal{W}^{11}_{\mbf{k}_1-\mbf{k}_i}
\mathcal{W}^{11}_{\mbf{k}_2-\mbf{k}_j}
\mathcal{W}^{11}_{\mbf{k}_3-\mbf{k}_l}
\mathcal{W}^{11}_{\mbf{k}_{ijl}-\mbf{k}_{123}}
\notag \\
+
\int \dfrac{d \mbf{k}^3_i}{(2\pi)^3}
&
\int \dfrac{d \mbf{k}^3_j}{(2\pi)^3}
P^{(0)}(\mbf{k}_i)P^{(0)}(\mbf{k}_j)
\mathcal{W}^{11}_{\mbf{k}_1-\mbf{k}_i}
\mathcal{W}^{11}_{\mbf{k}_2+\mbf{k}_i}
\mathcal{W}^{11}_{\mbf{k}_3-\mbf{k}_j}
\mathcal{W}^{11}_{\mbf{k}_j-\mbf{k}_{123}}
+3\,\mrm{cyc.}
\notag \\
+
\int\dfrac{d \mbf{k}^3}{(2\pi)^3}  
&
P^{(0)}(\mbf{k})
\mathcal{W}^{21}_{-\mbf{k}_2-\mbf{k}_3}
\mathcal{W}^{11}_{\mbf{k}_2-\mbf{k}}
\mathcal{W}^{11}_{\mbf{k}_3+\mbf{k}}
+5\,\mrm{cyc.}
\notag \\
+
\int\dfrac{d\mbf{k}_i^3}{(2\pi)^3}   
&
\int\dfrac{d\mbf{k}_j^3}{(2\pi)^3} 
B^{(0)}(\mbf{k}_i,\mbf{k}_j)
\mathcal{W}^{21}_{-\mbf{k}_{23}-\mbf{k}_i}
\mathcal{W}^{11}_{\mbf{k}_2-\mbf{k}_j}
\mathcal{W}^{11}_{\mbf{k}_3+\mbf{k}_{ij}}
+5\,\mrm{cyc.}
\notag \\
+
\int\dfrac{d \mbf{k}^3}{(2\pi)^3} 
&
P^{(0)}(\mbf{k})
\mathcal{W}^{31}_{-\mbf{k}_2-\mbf{k}}
\mathcal{W}^{11}_{\mbf{k}_2+\mbf{k}}
+3\,\mrm{cyc.}
+
(1+\alpha^3)\mathcal{I}_{41}
\Bigg\}\,
\,,
\end{align}

\noindent where the normalisation factors $\mathcal{I}_{ij}$ \cite{Gil-Marin:2014sta} and window function transforms $\mathcal{W}_{ij}$ are defined as

\begin{eqnarray}
\mathcal{I}_{ij}=\int d\mbf{r}^3w^i(\mbf{r})\an^j(\mbf{r})
\,,\quad\quad
\mathcal{W}^{ij}_\mbf{k}=\int d\mbf{r}^3 w(\mbf{r})^i \an(\mbf{r})^j
    e^{i\mbf{r}\mbf{k}}\,,
\end{eqnarray}

\noindent and here  we use the galaxy weights $w$ and tracers density number $\an$ to compute the above quantities. 
The quadrupoles are obtained from the above definitions through the substitution $F_0(\mbf{k})\longrightarrow F_2(\mbf{k})$.
Note that the shot-noise terms that do not have as argument the $k$-mode with respect to which the LOS angle is considered (i.e., the quantity $F_2(\mbf{k})$), vanish by definition because they are isotropic. 
In order to estimate the shot-noise terms using measured quantities measured directly from the data, the integrals in the above Equation \ref{eq:fkp_continuous_estimators} for power spectrum and bispectrum estimators can be approximated as described in Appendix \ref{sec:fkp_estimators_appendix} by

\begin{align}
\label{eq:text_PB_estimator}
\langle F_0(\mbf{k})F_0(\mbf{k})\rangle 
&=
\hat{P}^{(0)}(k) + 
(1+\alpha )\dfrac{\mathcal{I}_{21}}{\mathcal{I}_{22}}
\,,
\notag \\
\langle F_0(\mbf{k})F_2(\mbf{k})\rangle 
&=
\hat{P}^{(2)}(k)
\,,
\notag \\
\langle F_0(\mbf{k}_1)F_0(\mbf{k}_2)F_0(-\sum_{i=1}^2\mbf{k}_i)\rangle 
&=
\hat{B}^{(0)}(k_1,k_2,k_3)
+ \dfrac{\mathcal{I}_{32}}{\mathcal{I}_{33}}\left[\hat{P}^{(0)}(k_1) +2\,\mrm{cyc.}\right]
+
(1-\alpha^2)\dfrac{\mathcal{I}_{31}}{\mathcal{I}_{33}}
\,,
\notag \\
\langle F_2(\mbf{k}_1)F_0(\mbf{k}_2)F_0(-\sum_{i=1}^2\mbf{k}_i)\rangle 
&=
\hat{B}^{(200)}(k_1,k_2,k_3)
+ \dfrac{\mathcal{I}_{32}}{\mathcal{I}_{33}}\hat{P}^{(2)}(k_1)
\,,
\end{align}

\noindent where in the last line we follow  the notation of \cite{Gualdi:2020ymf} for the bispectrum multipoles and for simplicity we explicitly write the expression of only one of the three bispectrum multipoles.
For the i-trispectrum, Equation \ref{eq:fkp_continuous_estimators} can be also approximated as

\begin{align}
\label{eq:text_T_estimator}
\langle F_0(\mbf{k}_1)F_0(\mbf{k}_2)F_0(\mbf{k}_3)F_0(-\sum_{i=1}^3\mbf{k}_i)\rangle 
&=
\hat{\mathcal{T}}^{(0)}(k_1,k_2,k_3,k_4)
\notag \\
+
\dfrac{1}{N_{D_1}N_{D_2}}
\sum^{N_{D_1}}_i
\sum^{N_{D_2}}_j
\Bigg\{
\dfrac{\mathcal{I}_{43}}{\mathcal{I}_{44}}
\Big[
\hat{B}^{(0)}(k_1,k_2,D_{1,i}) &+ 
\hat{B}^{(0)}(k_1,k_3,|\mbf{k}_2+\mbf{k}_4|) + 
\hat{B}^{(0)}(k_1,k_4,D_{2,j}) 
\notag \\
+
\hat{B}^{(0)}(k_2,k_3,D_{2,j}) &+ 
\hat{B}^{(0)}(k_2,k_4,|\mbf{k}_1+\mbf{k}_3|) + 
\hat{B}^{(0)}(k_3,k_4,D_{1,i})
\Big]
\notag \\
+\dfrac{\mathcal{I}_{42}}{\mathcal{I}_{44}}
\Big[
\hat{P}^{(0)}(k_1)+\hat{P}^{(0)}(k_2)&+
\hat{P}^{(0)}(k_3)+
\hat{P}^{(0)}(k_4)
\notag \\ 
+\hat{P}^{(0)}(D_{1,i})+\hat{P}^{(0)}(D_{2,j})&+
\hat{P}^{(0)}(|\mbf{k}_1+\mbf{k}_3|)
\Big]
\Bigg\}
+
(1+\alpha^3)\dfrac{\mathcal{I}_{41}}{\mathcal{I}_{44}}
\,,
\notag \\
\langle F_2(\mbf{k}_1)F_0(\mbf{k}_2)F_0(\mbf{k}_3)F_0(-\sum_{i=1}^3\mbf{k}_i)\rangle 
&=
\hat{\mathcal{T}}^{(2000)}(k_1,k_2,k_3,k_4)
\notag \\
+
\dfrac{1}{N_{D_1}N_{D_2}}
\sum^{N_{D_1}}_i
\sum^{N_{D_2}}_j
\Bigg\{
\dfrac{\mathcal{I}_{43}}{\mathcal{I}_{44}}
\Big[
\hat{B}^{(200)}(k_1,k_2,D_{1,i}) &+ 
\hat{B}^{(200)}(k_1,k_3,|\mbf{k}_2+\mbf{k}_4|) + 
\hat{B}^{(200)}(k_1,k_4,D_{2,j}) 
\notag \\
+
\hat{B}^{(002)}(k_2,k_3,D_{2,j}) &+ 
\hat{B}^{(002)}(k_2,k_4,|\mbf{k}_1+\mbf{k}_3|) + 
\hat{B}^{(002)}(k_3,k_4,D_{1,i})
\Big]
\notag \\
+\dfrac{\mathcal{I}_{42}}{\mathcal{I}_{44}}
\Big[
\hat{P}^{(2)}(k_1)+\hat{P}^{(2)}(D_{1,i})&+
\hat{P}^{(2)}(D_{2,j})+\hat{P}^{(2)}(|\mbf{k}_1+\mbf{k}_3|)
\Big]
\Bigg\}\,.
\end{align}

\noindent 
Differently from the power spectrum and bispectrum cases, the quantity  we decide to consider and measure for 4pt correlator in Fourier space is not the trispectrum itself,  but the integrated trispectrum $\mathcal{T}$ \cite{Sefusatti:2004xz,Gualdi:2020eag}. This estimator naturally averages the signal of all skew-quadrilaterals having four sides $(k_1,k_2,k_3,k_4)$ with different diagonals $D_1$ and $D_2$ (or diagonal $D_1$ and folding angle $\psi$ around it as defined in \cite{Gualdi:2020eag}). $N_{D_1}$ and $N_{D_2}$ are indeed the number of possible diagonals $D_1$ and $D_2$ for a given set of four sides $(k_1,k_2,k_3,k_4)$. The multipoles expansion of $\mathcal{T}$ is described in \cite{Gualdi:2021yvq} and follows what done for $P^{(\ell)}$ and $B^{(\ell)}$, i.e., the multipoles of order $\ell>1$ are obtained by associating an $\ell$-order Legendre polynomial to one of the $k$'s and using its orientation with respect to the LOS.

In deriving the above estimators for the bispectrum and i-trispectrum, it is customary to employ measured quantities to account for the shot-noise term in order to avoid systematic errors in the theoretical modelling of the correction \cite{Gil-Marin:2014sta}, hence we approximate the integrals present in Equation \ref{eq:fkp_continuous_estimators} in the limit of the weights being of order unity, as described in Appendix \ref{sec:fkp_estimators_appendix} where  the estimators' full derivation can also be found. 
The validity  of this simplification is tested in Section \ref{sec:shot_noise_test}.

Figure \ref{fig:T_det_all_quads} shows the i-trispectrum monopole (first row) and quadrupoles (following rows) measured from BOSS DR12 CMASS NGC. 
The signal is shown as a function of all asymmetric skew-quadrilateral configurations (no configurations of the type $(k_a,k_a,k_a,k_a)$ or $(k_a,k_a,k_b,k_b)$) that can be found for the chosen $k$-range.
Being $\mathcal{T}$ an integrated quantity, each data point strictly speaking includes a family of skew-quadrilaterals, as explained in Section 2.1  of \cite{Gualdi:2020eag}. Here and hereafter we will loosely refer to such a set of skew-quadrilaterals as "configuration" or simply "quadrilateral". 

From the figure it is evident the compatibility in both shape and amplitude between the measurements from the BOSS data with the mean and scatter given by the ones from the Patchy Mocks. This justifies the usage of the Patchy mocks for both estimating the i-trispectrum covariance matrix and studying the relative signal detection.

Immediately by visual-inspection it appears that certain configurations i-trispectrum signal has noticeably larger amplitude than neighbouring ones. By comparison with our previous works \cite{Gualdi:2020eag,Gualdi:2021yvq} on simulations with periodic boundary conditions \cite{Villaescusa-Navarro:2019bje}, these quadrilaterals's signal-enhancement is a clear effect of the convolution between the physical signal and the survey window function. It is known that the survey mask induces mode-coupling. For the i-trispectrum this results into "pseudo-unconnected" terms of the 4pt correlator in Fourier space (see \cite{Gualdi:2020eag} for more details on the expansion in terms of connected and unconnected components). Unconnected terms should only appear for quadrilaterals with pairs of equal sides such as $(k_a,k_a,k_b,k_b)$; the mode-coupling induced by the window function produces a similar resonance between specific configurations even if the $k$-modes involved are different. 

Including these configurations in the analysis would strongly bias the results by returning a much stronger detection mainly due to the effect of the survey window coupling to the physical signal.

For a similar reason we exclude symmetric configurations which would need the subtraction of unconnected terms. These are expected from previous work on dark matter simulations \cite{Gualdi:2020eag} to have an amplitude at least two orders of magnitude larger than the connected term containing the physical signal. Therefore given that the measurements from galaxy catalogues are much more noisier than from dark matter particles ones (mainly due to the much lower objects number density), even a small relative error in estimating the unconnected terms would strongly affect the derived unconnected component.

\begin{figure}[tbp]
\centering 
\includegraphics[width=1.\textwidth]
{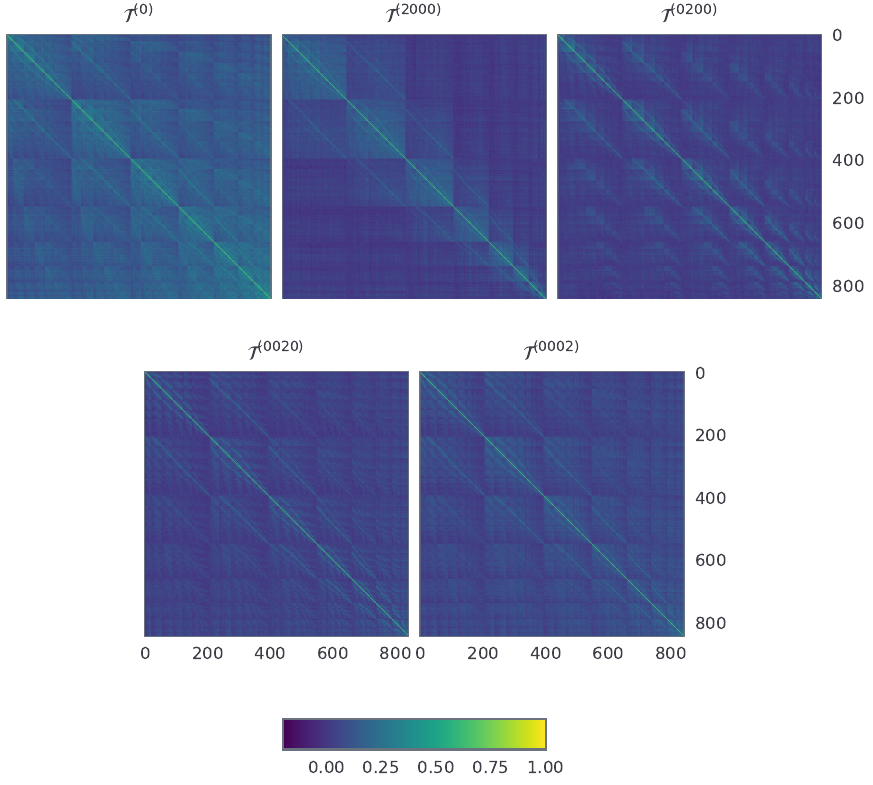}
\caption{\label{fig:T02_redcovs}
Reduced covariance matrices for the i-trispectrum multipoles estimated from 2048 Patchy Mocks realisations. Each reduced covariance matrix is computed as $\mrm{C}_{ij}/\sqrt{\mrm{C}_{ii}\mrm{C}_{jj}}$.
}
\end{figure}

In view of a simpler interpretation of the signal, an 
empirical a-posteriori procedure to exclude configurations whose signal is dominated by the window-induced mode coupling is presented in Appendix \ref{sec:remove_quads} together with showing the resulting data-vector in Figure \ref{fig:T_det_sel_quads}. In Figure \ref{fig:T02_redcovs} the reduced covariance matrices for each i-trispectrum multipole obtained from the 2048 Patchy Mocks are shown after the quadrilaterals selection step.
In Appendix \ref{sec:p_b_results} in Figure \ref{fig:PB_det} are reported the measurements for power spectrum and bispectrum multipoles.

\begin{figure}[tbp]
\centering 
\includegraphics[width=1.\textwidth]
{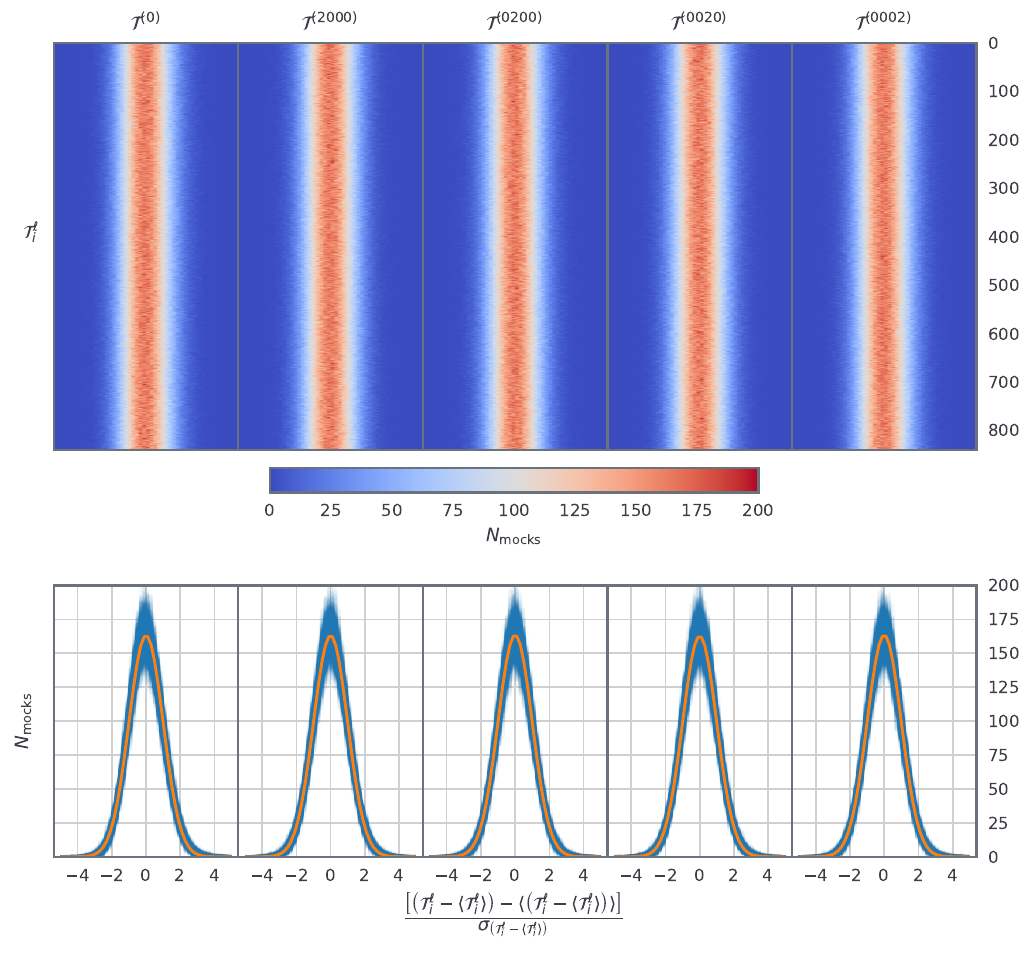}
\caption{\label{fig:gaussian_check_t}
Gaussianity test for the i-trispectrum data-vector monopole and quadrupoles. In the top panel each $i$-th row  corresponds to a i-trispectrum configuration,  the intensity map shows the histogram for the normalised scatter between the measure of $\mathcal{T}^\ell$ from each of the 2048 mock and the average $\langle\mathcal{T}^\ell\rangle$. The bottom panel shows the superposition of the normalised histograms for all the quadrilaterals (one blue line for each quadrilateral) while the orange line is a Gaussian distribution of the form $y(x)=A\times e^{-\frac{x^2}{2}}$ having by construction null median and unitary standard deviation, with $A$ being an amplitude set equal to the average of the maximum value of all the quadrilaterals curves (in blue). 
}
\end{figure}

\subsection{Signal Detection}
\label{sec:signal_detection}

The aim of this work is to show that there is a detectable i-trispectrum signal in both monopole and quadrupoles  measured from BOSS DR12 CMASS NGC data and that the signal is similar to the one measured from the survey galaxy mock catalogues. This second step is also essential to justify the i-trispectrum covariance matrix's estimation using the Patchy Mocks. To this aim, two quantities will be considered:

\begin{itemize}
    \item \textbf{signal-to-noise ratio (S/N)}: measures the presence of signal in the data and its strength with respect to cosmic variance;
    \item \textbf{$\mbf{\chi^2}$-test}: compares the summary statistics measured from the data with the average of the measurements from the Patchy mocks. This quantifies the similarity 
    between the statistical properties of the survey data and the galaxy mock catalogues.
\end{itemize}

\noindent Given a (measured) data-vector $\mbf{x}$ and the relative covariance matrix $\mrm{Cov}_\mbf{x}$, the signal-to-noise ratio ($\mrm{S/N}$) is computed as 

\begin{align}
\label{eq:ston}
    \mrm{S/N}= \sqrt{\mbf{x}^\intercal\, \mrm{Cov}_\mbf{x}^{-1}\,\mbf{x}}\,.
\end{align}

\noindent For the $\chi^2$-test we adopt

\begin{align}
    \label{eq:chi2_corr}
    \chi^2(\mbf{x}_i) = \left(\mbf{x}_i - 
    \langle\mbf{x}\rangle\right)^\intercal
    \mrm{Cov}^{-1}_\mbf{x}\left(\mbf{x}_i - 
    \langle\mbf{x}\rangle\right)\,,
\end{align}

\noindent where $\langle\mbf{x}\rangle$ is the average data-vector measured from a set of realisations. 
The bias induced by estimating the covariance matrices from a limited number of realisations \cite{Hartlap:2006kj,Sellentin:2016psv} has been accounted for by multiplying each inverse covariance by the appropriate Hartlap factor \cite{Hartlap:2006kj}, which is given by $h_\mrm{f}=(N_\mrm{mocks} - N_\mrm{dim} - 2)/(N_\mrm{mocks} - 1)$, where $N_\mrm{mocks}$ is the number of mock catalogues used to estimate the covariance while $N_\mrm{dim}$ is the dimension of the associated data-vector. We are aware that this correction is only approximated and that the correct treatment is presented in \cite{Sellentin:2016psv}. Given the number of simulations available and the size of the data vector,  and the direct comparison with the null hypothesis (see below), this approximation is sufficient to our purposes and of easy and immediate interpretation. 
For comparison we also construct  the null-hypothesis (N.H.)  case - no physical signal in the data - by generating 2048 fake data-vectors for power spectrum, bispectrum and i-trispectrum multipoles, centered around zero and with noise given by the diagonal elements of the relative covariance matrices estimated from the Patchy Mocks. In other words, the value for a given element $x_i$ of the data-vector $\mbf{x}$ will be generated from a Gaussian distribution with null mean and standard deviation $\sigma_i = \sqrt{\mrm{Cov}_\mbf{x}^{ii}}$. The covariance for the N.H. realisations is then computed together with the distribution's mean. The same analysis can then be performed both on the N.H. realisations, Patchy Mocks and BOSS data. This choice of N.H. is conservative in the sense that for the N.H. no correlation is assumed between different elements of the data-vectors, which instead is known to be present in the physical signal. For example we do not consider the additional correlation due to the effect of the survey window function on the N.H. data-vectors. Correlation means partial or full redundancy between different data-vectors elements and therefore is expected to reduce the $S/N$ with respect to the case when no correlation is present (diagonal covariance matrix, which is clearly not the case of the i-trispectrum multipoles covariances shown in Figure \ref{fig:T02_redcovs}).

The comparison between the $S/N$-distribution obtained from the Patchy Mocks (and the single values relative to the BOSS measurements) and the $S/N$-distribution given by the N.H. realisations defines the presence and significance of the physical signal in both synthetic and survey data. Assuming the N.H. realisations $S/N$-distribution to be well described by a Gaussian curve, the detection can be quantified in terms of $\sigma$-intervals from the null-hypothesis.

All the above quantities assume the data-vector $\mbf{x}$ to have a Gaussian distribution. This assumption must not hold in detail, but in Figure \ref{fig:gaussian_check_t} we illustrate how this assumption holds in practice in the case of the i-trispectrum. 
For each quadrilateral $i$  we compute the normalised distribution of the difference between the measurement of $\mathcal{T}^\ell_i$ for each mock $j$  and the average for all the 2048 mocks.

In the upper part of Figure \ref{fig:gaussian_check_t} the intensity map shows, for each multipole, the difference histogram obtained from all the catalogues, with a row for each quadrilateral configuration. The bottom part shows the superposition of all the quadrilaterals normalised difference w.r.t. the average distributions (blue lines); the orange line corresponds to  a normal distribution with amplitude given by the average maximum value of all the stacked curves.

It is clear that on average the data-vector has a Gaussian distribution for all the quadrilaterals. Therefore being a product of Gaussian curves, the data-vector's distribution is also a Gaussian curve. This allows us to use  and easily interpret the above estimators in Equations \ref{eq:ston} and \ref{eq:chi2_corr} for the signal's detection.

\begin{figure}[tbp]
\centering 
\includegraphics[width=1.\textwidth]
{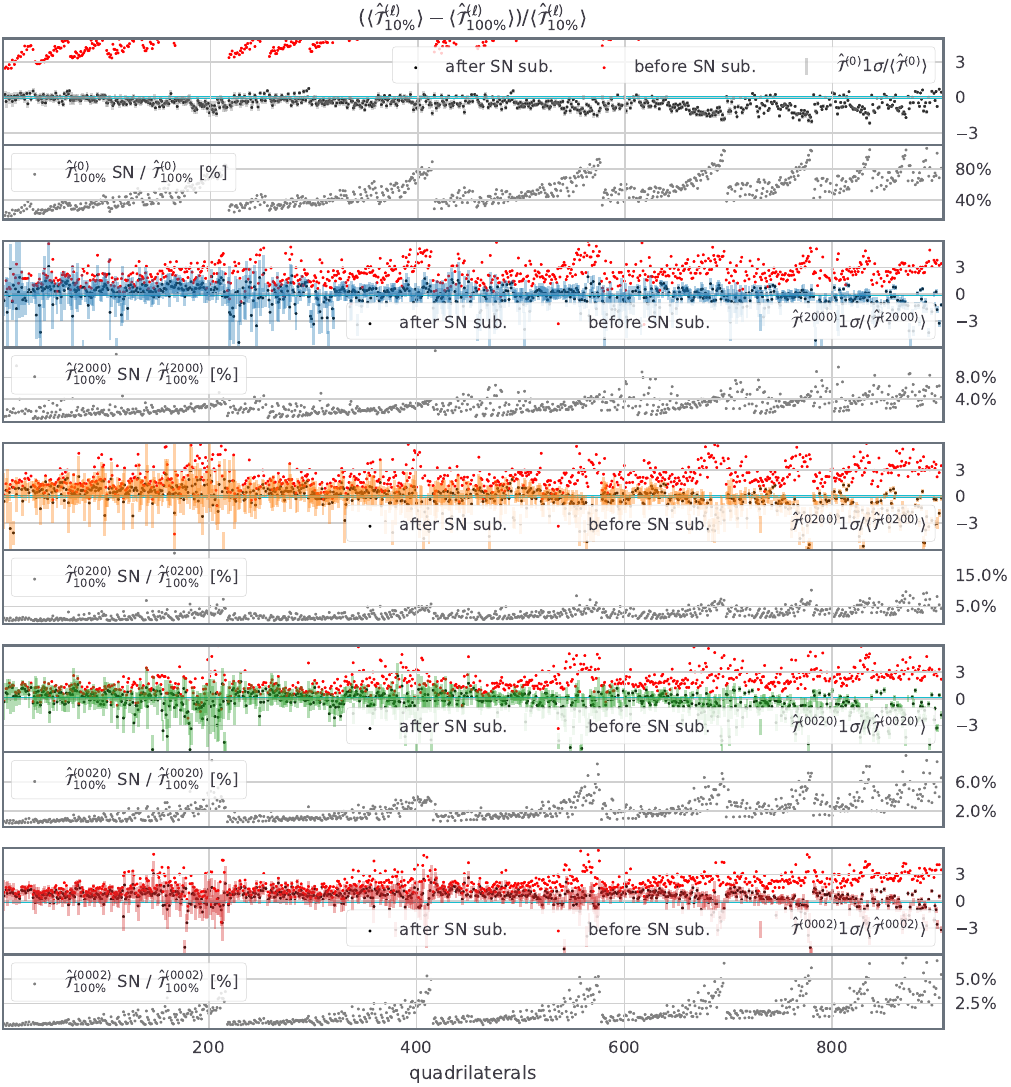}
\caption{\label{fig:T_shot_noise_test}
Shot-noise subtraction test for the i-trispectrum estimator (see Figure \ref{fig:PB_shot_noise} in Appendix \ref{sec:p_b_results} for the power spectrum and bispectrum equivalent check). The figure is divided vertically in 5 sections, one for the monopole and one for each quadrupole. Each section has two panels. The top panel displays  the relative difference between measurements performed on catalogues with different tracers density ($100\%$ and $10\%$ of the original catalogues) and are both shown before (red points) and after (black points) shot-noise subtraction.  The error bars are obtained from the standard deviation of the 2048 Patchy mocks.
 It is clear that the difference is compatible with zero after shot-noise subtraction for the majority of the quadrilaterals, contrary to what happens for the red points. In the bottom panel for each multipole, the ratio between the shot-noise correction and the statistic ${\cal T}^{(\ell)}$ (after shot-noise subtraction) is shown in the case of the full-density catalogue.
}
\end{figure}

\section{Results}
\label{sec:results}

For the different statistics we report measurements obtained with the following scale-cuts:

\begin{itemize}
    \item power spectrum: $k_\mrm{min}=0.02\,h/\mrm{Mpc}$ and $k_\mrm{max}=0.2\,h/\mrm{Mpc}$;
    \item bispectrum: $k_\mrm{min}=0.03\,h/\mrm{Mpc}$ and $k_\mrm{max}=0.15\,h/\mrm{Mpc}$;
    \item i-trispectrum: $k_\mrm{min}=0.03\,h/\mrm{Mpc}$ and $k_\mrm{max}=0.15\,h/\mrm{Mpc}$;
\end{itemize}

\noindent with a binning in $k$-space equal to $\Delta k=5\times k_\mrm{f}$ where $k_\mrm{f}=2\pi\,L_\mrm{box}$ is the fundamental frequency for a box of side $L_\mrm{box}=3500\, h^{-1}\mrm{Mpc}$. These settings yield 20 $k$-modes for the power spectrum, 435 triangles for the bispectrum and 906 (840 after the selection described in Appendix \ref{sec:remove_quads}) quadrilaterals for the i-trispectrum. 
For the i-trispectrum we only use quadrilaterals with $k_1\neq k_2\neq k_3\neq k_4$ to avoid configurations with unconnected terms \cite{Gualdi:2020eag}. Together with the implicit ordering ($k_1<k_2<k_3<k_4$) to avoid repetitions in the generation of quadrilateral configurations, the above choice implies that for example the quadrupole associated to $k_4$, $\mathcal{T}^{(0002)}$, will probe anisotropies associated to a smaller range of $k$-modes, mainly covering small-scales.
The above choice for the binning width in $k$-space is motivated as 
to have, for all the statistics, a final data-vector whose dimension is significantly smaller than the number of realisations used to estimate the covariance matrix and its inverse \cite{Hartlap:2006kj,Sellentin:2016psv}.  In the main text we focus on the i-trispectrum, 
the results relative to power spectrum and bispectrum are reported in Appendix \ref{sec:p_b_results}.

\subsection{Shot-Noise modelling and  subtraction}
\label{sec:shot_noise_test}

In Section \ref{sec:fkp_estimators} we have introduced an approximation in simplifying the estimators that allows one to model and subtract shot-noise using quantities directly measured from the data. Any inaccuracy introduced by this approximation is expected to be amplified if applied to an heavily sub-sampled mock galaxy distribution. 
Using the estimators defined in Section \ref{sec:fkp_estimators} we measure for 100 mocks the statistics in two cases: using all the galaxies for each catalogue (i.e,. a number density similar to that of the survey data) and only a randomly selected sub-sample corresponding to $10\%$ of the total. In this case the shot-noise signal is amplified by a factor of 10. 

Figure \ref{fig:T_shot_noise_test} shows the difference between the complete mock catalogues measurements and the sub-sampled  ones, before (red points) and after (black points) subtracting the shot-noise terms. The points represent the mean of 100 realisations while the errorbars correspond to the survey volume of one realisation.

The importance of properly removing the shot-noise is clearly evident for the i-trispectrum monopole: using measurements from a catalogue with ten times less tracers and without shot-noise subtraction would imply overestimating the signal at least by a factor of four, already at large scales. At small scales, mostly visible in the last quadrilaterals of Figure \ref{fig:T_shot_noise_test}, the shot-noise tends to be slightly overestimated. For the quadrupoles the shot-noise becomes more and more relevant as the modulo of the four $k$-modes making up the quadrilateral increases. In the case of the full-density catalogue, Figure \ref{fig:T_shot_noise_test} shows on alternate rows the percentage ratio between the shot-noise correction and the statistic (after shot-noise subtraction) for each $\mathcal{T}^{(\ell)}$. As expected, because of isotropy the impact of the shot-noise correction is much stronger for the monopole than for the i-trispectrum quadrupoles.

The same shot-noise convergence test for power spectrum and bispectrum is reported in Figure \ref{fig:PB_shot_noise} in Appendix \ref{sec:p_b_results}: the performance of the shot noise subtraction recipe for the trispectrum compares well with the performance for the power spectrum and bispectrum. 
Therefore we can conclude that our estimator properly accounts for the shot-noise term and that the approximations do not induce a significant bias in our analysis.

\begin{figure}[tbp]
\centering 
\includegraphics[width=1.\textwidth]
{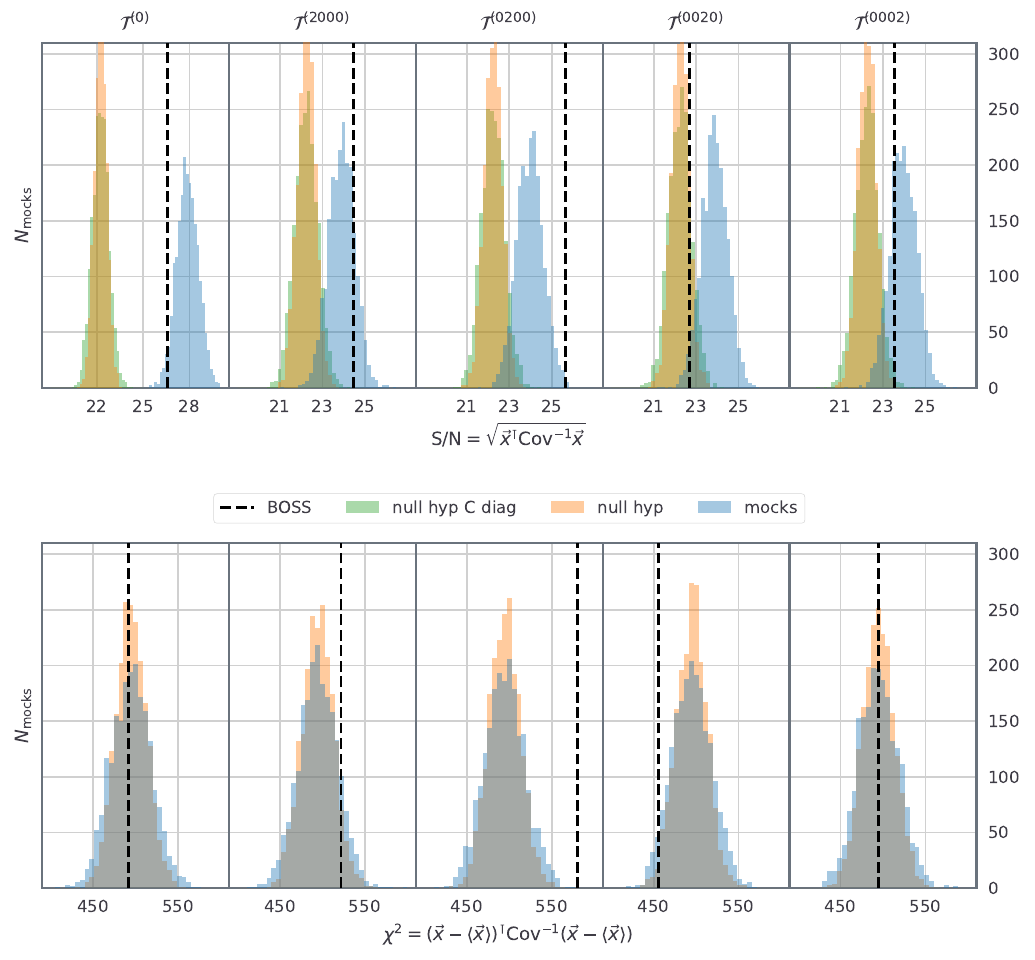}
\caption{\label{fig:T_ston_corr_chi2}
Detection statistics computed as described in Section \ref{sec:signal_detection} (Equations \ref{eq:ston} and \ref{eq:chi2_corr}) for the i-trispectrum monopole and quadrupoles. The orange histograms show the distributions obtained from the null-hypothesis (N.H.) realisations (no physical signal) while in blue are reported the distributions for the Patchy Mocks.
The black dashed lines are relative to the quantities computed for the BOSS DR12 NGC CMASS measurements. In the first row, the squared signal-to-noise ratio normalised by the number of degrees of freedom is reported. The bottom row displays the reduced $\chi^2$ obtained by comparing each realisation, including the data, with the mocks average. For the i-trispectrum monopole the difference between Patchy Mocks and N.H. distributions is very clear while  for the quadrupoles the distributions' tails partially overlap. The $S/N$ values for the Patchy Mocks distributions $(2.28\%,15.87\%,50\%,84.14\%,97.73\%)$ probability intervals together with $S/N$ results for the BOSS measurements are given in Table \ref{tab:ston_t}. In the same Table the distance from the N.H. in terms of $\sigma$-intervals is also reported.
}
\end{figure}

\subsection{Detection}
\label{sec:detection_results}

Figure \ref{fig:T_ston_corr_chi2} shows the statistical quantities described in Section \ref{sec:signal_detection} computed for i-trispectrum multipoles measurements from all 2048 Patchy mocks realisations and from BOSS DR12 CMASS NGC data, together with the N.H. realisations.

The first row displays the signal-to-noise ratio. This quantifies  presence and strength of i-trispectrum signal in both BOSS data and Patchy mocks. 
To assess the similarity between data and mocks on the second row the $\chi^2$-test between each Patchy Mocks measurement and the mocks' average is shown in blue. In both rows,  in orange the same is shown for the N.H. realizations. 
Monopole and quadrupoles approximately have same order of magnitude in terms of $S/N$, with monopole's $S/N$ being slightly larger. Notice that since the $(S/N)^2$ is equivalent to a $\chi^2$-test for a null theoretical model, it is expected that in the N.H. case the $(S/N)$-distribution peaks around the square-root of the data-vector's dimension (i.e., number of degrees of freedom). 

In the second row, the fact that the distribution for the (gravitational signal) mocks  and that for the N.H. realizations are so similar, indicates that there is nothing grossly unexpected in the $\chi^2$ distribution of the i-trispectrum in the mocks.

The dashed black line, representing the result of the measurement on BOSS data, falls well within the distribution given by the results for the 2048 Patchy mocks' histogram. This is confirmed by the $\chi^2$-test, proving that the i-trispectrum signal detected from data is compatible with the one present in the galaxy mock catalogues. In Table \ref{tab:ston_t} $S/N$ values corresponding to $(2.28\%,15.87\%,50\%,84.14\%,97\%)$ probability intervals of the Patchy mocks distribution shown in Figure \ref{fig:T_ston_corr_chi2} are reported together with the values obtained for BOSS data measurements. The detection significance in terms of $\sigma$-intervals from the N.H. distribution is also given in Table \ref{tab:ston_t} for the same probability intervals of the Patchy Mocks $S/N$ distribution and for the measurements on BOSS data. For the latter we find a detection significance for the i-trispectrum multipoles of $(10.4,5.2,8.3,1.1,3.1)$ $\sigma$’s away from the N.H. $S/N$ distribution.

In Appendix \ref{sec:p_b_results} Figure \ref{fig:PB_ston_corr_chi2} and Table \ref{tab:ston_pb} the corresponding results for power spectrum and bispectrum multipoles are reported. While the detection significance for $P^{(0,2)}$ and $B^{(0)}$ is significantly higher than for $\mathcal{T}^{(0)}$, it is interesting to note that the distance from the null-hypothesis for $B^{(2)}$ and $\mathcal{T}^{(2)}$ is similar.

\begin{figure}[tbp]
\centering 
\includegraphics[width=1.\textwidth]
{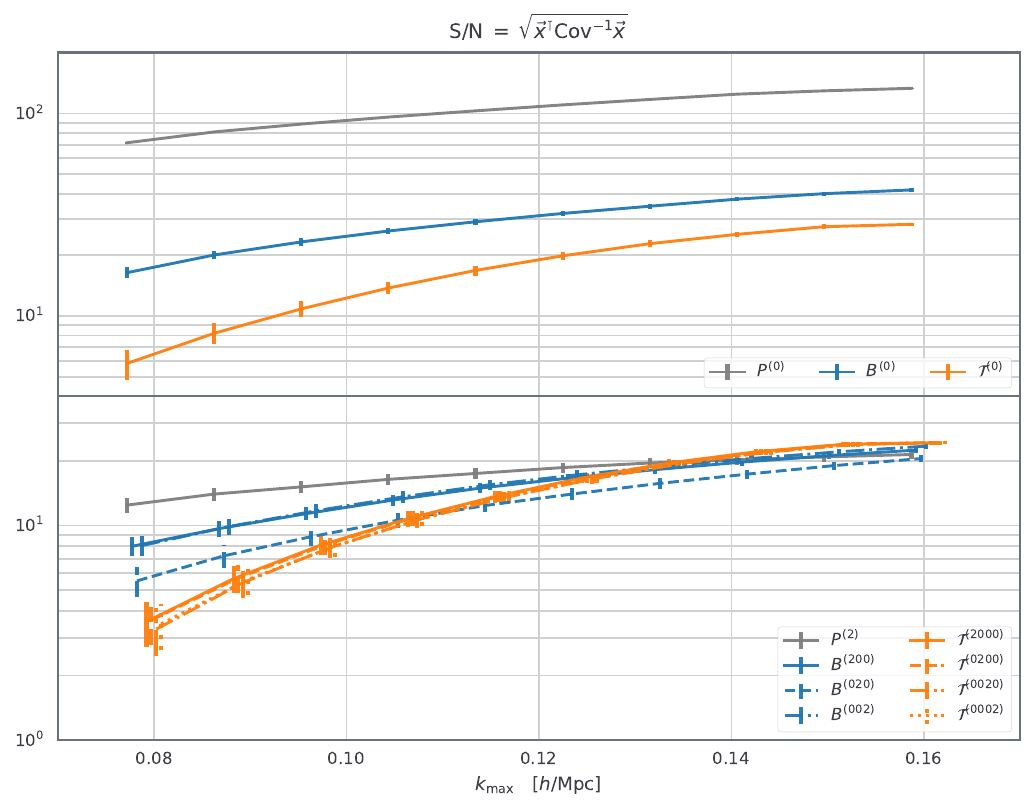}
\caption{\label{fig:signal_to_noise_PBT02}
Using Equation \ref{eq:ston} the signal-to-noise ratio ($S/N$) as a function of the maximum wave-number $k_\mrm{max}$ is computed for both monopole (upper panel) and quadrupole/s (lower panel) of power spectrum, bispectrum and i-trispectrum statistics measured from each of the 2048 Patchy mocks. The bispectrum and i-trispectrum quadrupoles points are artificially shifted on the $x$-axis to improve the clarity of the plot, with the $S/N$ corresponding to the same $k_\mrm{max}$ value. 
}
\end{figure}

In Figure \ref{fig:signal_to_noise_PBT02} a summary of the signal-to-noise ratio for both monopoles and quadrupoles for power spectrum, bispectrum and i-trispectrum is shown as a function of the maximum considered wave-number $k_\mrm{max}$. Even if lower, the i-trispectrum monopole's $S/N$ is of the same order of magnitude of the bispectrum monopole, with the gap between the two decreasing as $k_\mrm{max}$ increases.
Note that the i-trispectrum quadrupoles' $S/N$ reaches the same value of  power spectrum and bispectrum ones, supporting the claim  made in \cite{Gualdi:2021yvq} regarding $\mathcal{T}^{(2)}$'s potential to probe significant additional information with respect to $P^{(0,2)}$, $B^{(0,2)}$ and $\mathcal{T}^{(0)}$.

\renewcommand{\arraystretch}{2.}
\begin{table}[tbp]
\centering
\begin{tabular}{c|c|c|c|c|c|c|}
\cline{2-7}
 & \multicolumn{6}{c|}{$S/N$ and $(\Delta \sigma)$-distance } \\
\cline{2-7}
& 2.28 $\%$ & 15.87 $\%$ & 50 $\%$ & 84.14 $\%$ & 97.73 $\%$ & DATA\\
\hline
$\mathcal{T}^{(0)}$      & \textbf{26.4 (10.1)} & \textbf{27.1 (11.8)} & 27.9 (13.5)         & 28.6 (15.3)         & 29.3 (16.9)         & \textbf{26.6 (10.4)} \\
$\mathcal{T}^{(2000)}$   & 22.7 (1.0)           & 23.3 (2.5)           & \textbf{23.9 (4.0)} & \textbf{24.5 (5.4)} & 25.1 (6.8)          & \textbf{24.4 (5.2)} \\
$\mathcal{T}^{(0200)}$   & 22.7 (1.2)           & 23.3 (2.7)           & 24.0 (4.2)          & 24.6 (5.7)          & \textbf{25.2 (7.3)} & \textbf{25.7 (8.3)} \\
$\mathcal{T}^{(0020)}$   & \textbf{22.6 (0.9)}  & \textbf{23.3 (2.5)}  & 23.9 (3.9)          & 24.5 (5.4)          & 25.1 (6.8)          & \textbf{22.7 (1.1)} \\
$\mathcal{T}^{(0002)}$   & 22.7 (1.1)           & \textbf{23.3 (2.6)}  & \textbf{23.9 (4.0)} & 24.6 (5.6)          & 25.2 (7.0)          & \textbf{23.5 (3.1)} \\
\hline
\end{tabular}
\caption{\label{tab:ston_t}
$S/N$ values corresponding to the first row in Figure \ref{fig:T_ston_corr_chi2} for the different i-trispectrum multipoles. From left to right the columns refer to the $S/N$ for the $(2.28\%,15.87\%,50\%,84.14\%,97.73\%)$ probability intervals of the Patchy mocks $\mrm{S/N}$-distribution displayed in Figure \ref{fig:T_ston_corr_chi2}. Between brackets is reported the distance between $S/N$-values and the peak of $S/N$-distribution for the null hypothesis (N.H.) in terms of $\sigma$-intervals. The last column reports the corresponding $S/N$ (and $\sigma$-intervals distance) for the measurements on BOSS DR12 CMASS NGC data. Beside the data values, in bold font are highlighted the two closest probability intervals containing the data's results. The distance from the N.H. is much larger for the i-trispectrum monopole than for the quadrupoles. The same quantities for power spectrum and bispectrum multipoles are given in Table \ref{tab:ston_pb} where in particular the bispectrum quadrupoles have similar detection significance to the i-trispectrum ones. 
}
\end{table}

\section{Conclusions}
\label{sec:conclusions}

For the first time we report a detection of the i-trispectrum's monopole and quadrupoles signal from BOSS CMASS NGC DR12 data. 
To achieve this, the FKP estimators formalism \cite{Feldman:1993ky} was extended to the 4pt level to measure statistics from an observed volume with a non-regular survey geometry (without periodic boundary conditions). The resulting expression is reported in Equation \ref{eq:text_T_estimator}.
We then measure from BOSS data and 2048 realisations of the Patchy Mocks \cite{Kitaura:2015uqa,Rodriguez-Torres:2015vqa} the monopole and quadrupole terms for power spectrum, bispectrum and i-trispectrum.

The ability of the proposed estimator to account for the shot-noise is tested in Section \ref{sec:shot_noise_test} by measuring the statistics from 100 realisations of the Patchy Mocks both at full density and randomly subsampled down to $10\%$ of the original density. The difference between the two measurements is shown to be consistent with zero after shot-noise subtraction in Figure \ref{fig:T_shot_noise_test} (Figure \ref{fig:PB_shot_noise} for the same test for power spectrum and bispectrum).

The i-trispectrum data-vector effective Gaussian behaviour is shown  in Figure \ref{fig:gaussian_check_t}:  the normalised distribution for each quadrilateral signal is extracted from the 2048 mock measurements. 

As a reference for the null hypothesis (N.H.) - absence of a physical signal - 2048 artificial realisations were produced for each statistics generating each data-vector's element from a normal distribution with zero mean and standard deviation derived from the corresponding covariance matrix (estimated from Patchy Mocks) diagonal element. 

By computing the signal-to-noise ratio ($S/N$) as in Equation \ref{eq:ston} for both N.H. realisations and Patchy Mocks / BOSS measurements we can both quantify 

\begin{itemize}
    \item the absolute $S/N$ in BOSS data, which for the i-trispectrum monopole and quadrupoles ($\mathcal{T}^{(\ell)}$) results to be $(26.6,24.4,25.7,22.7,23.5)$, while the corresponding numbers for the null hypothesis $S/N$-distributions medians are $(22.2,22.2, 22.3, 22.2, 22.2)$.
    \item the distance between N.H. $S/N$-distribution and the data's $S/N$: the detection's significance expressed in terms of $\sigma$-intervals is $(10.4,5.2,8.3,1.1,3.1)$ $\sigma$'s .
\end{itemize}

\noindent These results are summarised in Figure \ref{fig:T_ston_corr_chi2} and Table \ref{tab:ston_t}.
Table \ref{tab:ston_t} also reports the $S/N$ for the $(2.28\%,15.87\%,50\%,84.14\%,97.73\%)$ probability intervals of the Patchy Mocks $S/N$-distribution and the relative distance from the N.H. $S/N$-distribution, always in terms of $\sigma$-intervals. Equivalent results for power spectrum and bispectrum multipoles are reported in Appendix \ref{sec:p_b_results} in Figure \ref{fig:PB_ston_corr_chi2} and Table \ref{tab:ston_pb}. 
The detection significance of $\mathcal{T}^{(0)}$ is much larger than the quadrupoles' ones. Nevertheless $\mathcal{T}^{(2)}$'s average distance from the N.H. is similar to the ones for the bispectrum quadrupoles ($B^{(2)}$).

The data's signal-to-noise ratio values for $\mathcal{T}^{(\ell)}$ always fall within the probability distribution mocks' $S/N$ values.
This further confirms the similarity between the physical signal measured from Patchy Mocks and BOSS data. This compatibility is also supported by the $\chi^2$-test (Equation \ref{eq:chi2_corr}) checking the compatibility with the mocks' average, which is also shown in the bottom row of Figure \ref{fig:T_ston_corr_chi2}.
To summarise, the i-trispectrum physical signal is present and detectable from BOSS data. Its analysis can be performed by employing the Patchy Mocks which also have an i-trispectrum signal compatible with the one from the data.

Figure \ref{fig:signal_to_noise_PBT02} compares the three statistics monopoles and quadrupoles $S/N$ as a function of $k_\mrm{max}$. In the monopoles' panel the power spectrum's $S/N$ is one order of magnitude larger than bispectrum and i-trispectrum ones with the difference remaining constant across the the $k_\mrm{max}$'s range. The trend changes in the quadrupoles' case: the signal-to-noise ratio for $\mathcal{T}^{(2)}$ is the lowest for the minimum $k_\mrm{max}$ but grows faster than $P^{(2)}$ and $B^{(2)}$ ones until becoming the highest at smaller scales.
This confirms using spectroscopic galaxy surveys data what we observed and proved using dark matter simulations in \cite{Gualdi:2021yvq}: $\mathcal{T}^{(\ell)}$ contains additional signal / information with respect to power spectrum and bispectrum and must be employed to fully exploit current and future clustering data-sets.

We expect that the i-trispectrum monopole and quadrupoles signal will be significantly stronger in incoming clustering datasets such as the one being currently produced by DESI and PFS surveys because of the much larger observed volumes.

The next steps towards the goal of constraining cosmological parameters using the i-trispectrum consist in studying the survey window effect on the signal and in extending the theoretical modelling to include more non-linear scales. Another important ingredient will be the data-vector's compression: given the $840$ available quadrilaterals, with only 2048 mock catalogues,  in this paper we couldn't  consider i-trispectrum's monopole and quadrupoles jointly. Hence, to fully exploit the i-trispectrum potential \cite{Gualdi:2021yvq} it will be then necessary to extend  current optimal compression techniques \cite{Gualdi:2019sfc} to the 4pt-correlation level.


\acknowledgments 
D.G. thanks H\'ector Gil-Mar\'in for the invaluable discussions and the IT team at ICCUB for the help with the Aganice cluster. 
L.V. and D.G. acknowledge support of European Union's Horizon 2020 research and innovation programme ERC (BePreSysE, grant agreement 725327). Funding for this work was partially provided by project PGC2018-098866- B-I00 MCIN/AEI/10.13039/501100011033 y FEDER “Una manera de hacer Europa”, and the “Center of Excellence Maria de Maeztu 2020-2023” award to the ICCUB (CEX2019-000918-M funded by MCIN/AEI/10.13039/501100011033)

\appendix

\section{FKP estimator}
\label{sec:fkp_estimators_appendix}
In this appendix we present the power spectrum, bispectrum and i-trispectrum estimators expressions in terms of the Feldman-Kaiser-Peacock (FKP) formalism \cite{Feldman:1993ky}. Below the well-known derivation for power spectrum and bispectrum is summarised in order to later introduce the computation of the i-trispectrum estimator.

\subsection{Power Spectrum}

Starting from the definition introduced by Peebles \cite{Peebles1980} of the probability of having a galaxy inside a volume element $\delta V$

\begin{eqnarray}
P = \delta V \Bar{n}\pos(1+f\pos)\,,
\end{eqnarray}

\noindent where $\Bar{n}\pos$ is the expected mean spatial density of galaxies given the angular and luminosity selection criteria and where the subscript indicates a dependence on the radial position vector $\mathbf{r}$.
The lowest order statistic we want to estimate is the power spectrum, defined as the Fourier transform of the 2pt correlation function $\xi$,

\begin{eqnarray}
 P\mom \equiv P_k= \int d\mbf{r}^3\xi\pos e^{i\mbf{kr}}
 \quad \mrm{where} 
 \quad 
 \xi\pos\equiv\xi_r=\langle f\pos f\ppos\rangle\,.
\end{eqnarray}

\noindent We use the Fourier convention

\begin{eqnarray}
 f\mom=\int d \mbf{r}^3f\pos e^{i\mbf{kr}}
 \quad
 \mrm{and}
 \quad
 f\pos=\dfrac{1}{(2\pi)^3}\int d \mbf{k}^3 f\mom e^{-i\mbf{kr}}\,.
\end{eqnarray}

\noindent Assuming $f\pos=0$ outside a very large volume $V$ we then have that

\begin{eqnarray}
 \dfrac{1}{V}\langle f\mom f\mom^*\rangle = \int d\mbf{r}^3\langle f\pos f\ppos\rangle e^{i\mbf{kr}}=P_k\,.
\end{eqnarray}

\noindent From the galaxy catalogue and a synthetic (s) one with identical radial and angular selection function we define the quantity

\begin{eqnarray}
\label{eq:F_fkp}
 F\pos=\dfrac
 {w\pos\left[n^\mrm{g}\pos-\alpha n\pos^\mrm{s}\right]}
 {\left[\int d\mbf{r}^3\bar{n}\pos^2w\pos^2\right]^{\frac{1}{2}}}
 \quad
 \mrm{where}
 \quad
 n\pos\gal=\sum_i \delta_D\left(\mbf{r}-\mbf{r}_i\right)\,,
\end{eqnarray}

\noindent where $w\pos$ is the weighting function evaluated at the position $\mbf{r}$, while $\alpha=N_\mrm{gal}/N_\mrm{s}$. Taking its Fourier transform, squaring it and computing the expectation value 

\begin{align}
\label{eq:pk_estimator}
    \langle F\mom F\mom^*\rangle=
    \dfrac{
    \int d\mbf{r}_a^3\int d\mbf{r}_b^3
    \langle
    w\ra w\rb
    \left[n\ra\gal-\alpha n\ra\rnd\right]
    \left[n\rb\gal-\alpha n\rb\rnd\right]
    \rangle
    e^{i\mom(\mbf{r}_a-\mbf{r}_b)}}
    {\int d\mbf{r}_a^3\bar{n}\ra^2w\ra^2}\,.
\end{align}

\noindent In order to simplify the terms appearing from the above product in the numerator we need to use Peebles's result for infinitesimal volumes where the occupation number's possible values are just $n=0,1$ \cite{Peebles1980}:

\begin{align}
    \langle n_a n_b\rangle=
    \begin{cases}
        \bar{n}\ra\bar{n}\rb \delta V^2\left[1+\xi\rab\right]\quad&\mrm{if}\quad a \neq b\\
        \langle n_a^2\rangle=\langle n_a\rangle=\bar{n}\ra\delta V \quad&\mrm{if}\quad a = b\,. \\
    \end{cases}
\end{align}

\noindent Then proceeding as in \cite{Feldman:1993ky} one has that for both galaxies and synthetic objects the following relations hold:

\begin{align}
    \langle n\gal\ra n\gal\rb\rangle &= \an\ra\an\rb\left[1+\xi\rab\right]+\an\ra\DD\rab
    \notag\\
    \langle n\rnd\ra n\rnd\rb\rangle &= \alpha^{-2}\an\ra\an\rb + \alpha^{-1}\an\ra\DD\rab
    \notag\\
    \langle n\gal\ra n\rnd\rb\rangle &= \alpha^{-1}\an\ra\an\rb,
\end{align}

\noindent using the 2pt correlation function definition given in \cite{Peebles1980}.
Using the above correlators, equation \ref{eq:pk_estimator} can be simplified to

\begin{align}
    \langle F\mom F\mom^*\rangle = 
    \dfrac{\int d\mbf{r}_a^3d\mbf{r}_b^3w\ra w\rb e^{i\mbf{k}(\mbf{r}_a-\mbf{r}_b)}
    \left[ n\ra n\rb \xi\rab + (1+\alpha)n\ra\DD\rab\right]}
    {\int d\mbf{r}_a^3\bar{n}\ra^2w\ra^2} \,.
    \label{eq:A9}
\end{align}

\noindent Recalling the relation between 2pt correlation function and power spectrum

\begin{align}
    \xi\rab = \dfrac{1}{(2\pi)^3}\int d \mbf{k}^3 e^{-i\mbf{k}(\mbf{r}_a - \mbf{r}_b)}P(k)\,,
\end{align}

\noindent  equation \ref{eq:A9} simplifies to

\begin{align}
\label{eq:pk_befassumption}
    \langle F\mom F\mom^*\rangle &= \dfrac{\int \dfrac{d\mbf{q}^3}{(2\pi)^3}P(q)
    \int d\mbf{r}_a^3w\ra\an\ra e^{i\mbf{r}_a(\mbf{k}-\mbf{q})}
    \int d\mbf{r}_b^3w\rb\an\rb e^{-i\mbf{r}_b(\mbf{k}-\mbf{q})}
    + (1+\alpha )\int d\mbf{r}_a^3w^2\ra\an\ra}
    {\int d\mbf{r}_a^3\bar{n}\ra^2w\ra^2}
    \notag \\
    &=
    \mathcal{I}_{22}^{-1}\int \dfrac{d\mbf{q}^3}{(2\pi)^3}P(q)
    \mathcal{W}(\mbf{k}-\mbf{q})\mathcal{W}^*(\mbf{k}-\mbf{q})
    \quad+\quad (1+\alpha ) \mathcal{I}_{22}^{-1}\mathcal{I}_{21}.
\end{align}

\noindent In the above equation a renormalising quantity $\mathcal{I}_{ij}$ together with the window function $\mathcal{W}^{ij}$ in Fourier space were introduced 
:

\begin{align}
\label{eq:renormalising_fac}
    \mathcal{I}_{ij} = \int d\mbf{r}_a^3w^i\ra\an^j\ra\,,
\quad\quad
    \mathcal{W}^{ij}\left(\mbf{k}\right)=
    \int d\mbf{r}^3 w\pos^i \an\pos^j
    e^{i\mbf{r}\mbf{k}}\,.
\end{align}

\noindent In equation \ref{eq:pk_befassumption} the second term corresponds to the shot-noise and  in the first term the power spectrum is convolved with the window function $\mathcal{W}$.


\subsection{Bispectrum}

For the three-point correlation function, in the case of an infinitesimal volume where the number of objects inside a cell can only be $n=0,1$, the possibilities for the correlator are:

\begin{align}
    \langle n_a n_b n_c\rangle=
    \begin{cases}
        \bar{n}\ra\bar{n}\rb\bar{n}\rc \delta V^3
        \left[1+\xi(\mbf{r}_{ab}) + \xi(\mbf{r}_{ac}) + \xi(\mbf{r}_{bc}) + \zeta(\mbf{r}_{ab},\mbf{r}_{ac})\right]\quad
        &\mrm{if}\quad a \neq b\neq c\\
        \langle n_a^2 n_c\rangle=\langle n_a n_c\rangle=\bar{n}\ra\bar{n}\rc\delta V^2
        \left[1+\xi(\mbf{r}_{ac})\right]\quad
        &\mrm{if}\quad a = b \neq c\,, \\
        \langle n_c^2 n_b\rangle=\langle n_c n_b\rangle=\bar{n}\rc\bar{n}\rb\delta V^2
        \left[1+\xi(\mbf{r}_{bc})\right]\quad
        &\mrm{if}\quad a = c \neq b\,, \\
        \langle n_b^2 n_a\rangle=\langle n_a n_b\rangle=\bar{n}\ra\bar{n}\rb\delta V^2
        \left[1+\xi(\mbf{r}_{ab})\right]\quad
        &\mrm{if}\quad b = c \neq a\,, \\
        \langle n_a^3\rangle=\langle n_a\rangle=\bar{n}\ra\delta V \quad
        &\mrm{if}\quad a = b = c\,. \\
    \end{cases}
\end{align}

\noindent Then, proceeding again as in \cite{Feldman:1993ky},

\begin{align}
    &\langle \int d\mbf{r}_a^3d\mbf{r}_b^3d\mbf{r}_c^3\,g(\mbf{r}_a,\mbf{r}_b,\mbf{r}_c)n\ra n\rb n\rc\rangle  = \sum_{i\neq j\neq k} \an_i\an_j\an_k\delta V^3\left[1 + \xi_{ij} + \xi_{ik} + \xi_{jk} + \zeta_{ijk}\right]\,g(\mbf{r}_i,\mbf{r}_j,\mbf{r}_k)
    \notag \\
    &+
    \sum_{i=j\neq k }\an_i \an_k \delta V^2\left[1 + \xi_{ik}\right]\,g(\mbf{r}_i,\mbf{r}_i,\mbf{r}_k)\quad+\quad2\,\mrm{cyc.}\quad+
    \sum_{i=j=k}\an_i\delta V
    \notag \\
    &=
    \int d\mbf{r}_a^3d\mbf{r}_b^3d\mbf{r}_c^3\,g(\mbf{r}_a,\mbf{r}_b,\mbf{r}_c)
    \Bigg\{
    \an\ra\an\rb\an\rc\left[1+ \xi\rab + \xi\rac + \xi\rbc + \zeta\rabc\right]
    \notag \\
    &+
    \an\ra\an\rc\left[1+ \xi\rac\right]\delta^D\rab\quad+\quad2\,\mrm{cyc.}\quad+
    \an\ra\delta^D\rab\delta^D\rac
    \Bigg\}\,,
\end{align}

\noindent from which the expansion for the correlator of both galaxies and synthetic objects can be found

\begin{align}
\label{eq:3pt_terms_exp}
    \langle n\gal\ra n\gal\rb n\gal\rc\rangle &= 
    \an\ra\an\rb\an\rc\left[1+ \xi\rab + \xi\rac + \xi\rbc + \zeta\rabc\right]
    +
    \an\ra\an\rc\left[1+ \xi\rac\right]\delta^D\rab+2\,\mrm{cyc.}+
    \an\ra\delta^D\rab\delta^D\rac
    \notag \\
    \notag \\
    \langle n\gal\ra n\gal\rb n\rnd\rc\rangle 
    &= \alpha^{-1}\{\an\ra\an\rb\an\rc\left[1+ \xi\rab\right] + \an\ra\an\rc\delta^D\rab\}
    \notag \\
    \notag \\
    \langle n\gal\ra n\rnd\rb n\rnd\rc\rangle 
    &= \alpha^{-2}\{\an\ra\an\rb\an\rc + \alpha\an\ra\an\rb\delta^D\rbc\}
    \notag \\
    \notag \\
    \langle n\rnd\ra n\rnd\rb n\rnd\rc\rangle 
    &= \alpha^{-3}\{\an\ra\an\rb\an\rc 
    + \alpha\left[\an\ra\an\rb\delta^D\rac + \an\ra\an\rc\delta^D\rbc + \an\rb\an\rc\delta^D\rab\right]
    + \alpha^2\an\ra\delta^D\rab\delta^D\rac\}\,.
\end{align}

\noindent In \cite{Gil-Marin:2014sta} the field $F$ was defined as $F_i(\mbf{r})\equiv w_\mr{FKP}(\mbf{r})\lambda_i\left[w_\mrm{c} n_{\mbf{r}} - \alpha n\rnd_{\mbf{r}}\right]$ with $\lambda\equiv \mathcal{I}_i^{-\frac{1}{i}}$ and $\mathcal{I}_i\equiv\int d\mbf{r}^3w_\mrm{FKP}^i(\mbf{r})\langle n w_\mrm{c}\rangle^i(\mbf{r})$.
For the bispectrum $i$ was set $i=3$. Following this convention we can compute the three-point correlator in Fourier space with the specification that $\mbf{k}_3 = - \mbf{k}_1-\mbf{k}_2$:

\begin{align}
\label{eq:3ptest}
    &\langle F(\mbf{k}_1)F(\mbf{k}_2)F(\mbf{k}_3)\rangle = 
    \mathcal{I}^{-1}_{33}\int d\mbf{r}_a^3d\mbf{r}_b^3d\mbf{r}_c^3 w\ra w\rb w\rc
      \notag \\
    &\times
    \langle\left[n\ra - \alpha n\rnd\ra\right]\left[n\rb - \alpha n\rnd\rb\right]
    \left[n\rc - \alpha n\rnd\rc\right]\rangle e^{i(\mbf{k}_1\mbf{r}_a+\mbf{k}_2\mbf{r}_b+\mbf{k}_3\mbf{r}_c)}
    \notag \\
    &=
        \mathcal{I}^{-1}_{33}\times\Bigg\{
    \int d\mbf{r}_a^3d\mbf{r}_b^3d\mbf{r}_c^3 w\ra w\rb w\rc
    \an\ra\an\rb\an\rc\zeta\rabc
    e^{i(\mbf{k}_1\mbf{r}_a+\mbf{k}_2\mbf{r}_b-(\mbf{k}_1+\mbf{k}_2)\mbf{r}_c)}
    \notag\\
    &+
    \int d\mbf{r}_a^3d\mbf{r}_b^3 w\ra w\rb^2
    \an\ra\an\rb\xi\rab
    e^{i\mbf{k}_1\left(\mbf{r}_a-\mbf{r}_b\right)}
    +
    \int d\mbf{r}_b^3d\mbf{r}_c^3 w\rb w\rc^2
    \an\rb\an\rc\xi\rbc
    e^{i\mbf{k}_2\left(\mbf{r}_b-\mbf{r}_c\right)}
    \notag\\
    &+
    \int d\mbf{r}_a^3d\mbf{r}_c^3 w\ra^2w\rc
    \an\ra\an\rc\xi\rac
    e^{i\left(\mbf{k}_1+\mbf{k}_2\right)\left(\mbf{r}_a-\mbf{r}_c\right)}
    +
    (1-\alpha^2)
    \int d\mbf{r}_a^3 w\ra^3\an\ra
    \Bigg\}\,.
    \notag \\
\end{align}

\noindent Before proceeding let's recall the relation between 3pt correlation function and bispectrum in terms of inverse Fourier transform:

\begin{align}
    \zeta\rabc=\zeta(\mbf{r}_a-\mbf{r}_c,\mbf{r}_b-\mbf{r}_c)=
    \int\dfrac{d\mbf{k}_i^3d\mbf{k}_j^3}{(2\pi)^6}
    B(\mbf{k}_i,\mbf{k}_j)
    e^{-i\left[\mbf{k}_i(\mbf{r}_a-\mbf{r}_c)+\mbf{k}_j(\mbf{r}_b-\mbf{r}_c)\right]}\,.
\end{align}

\noindent Expanding both 2pt and 3pt correlation functions in Equation \ref{eq:3ptest} in terms of power spectra and bispectrum one obtains:

\begin{align}
    &\langle F(\mbf{k}_1)F(\mbf{k}_2)F(-\mbf{k}_1-\mbf{k}_2)\rangle 
    =
\mathcal{I}^{-1}_{33}\times\Bigg\{
    \int\dfrac{d\mbf{k}_i^3d\mbf{k}_j^3}{(2\pi)^6}
    B(\mbf{k}_i,\mbf{k}_j)
    \notag\\
    &\times
    \int d\mbf{r}_a^3w\ra\an\ra e^{i\mbf{r}_a(\mbf{k}_1-\mbf{k}_j)}
    \int d\mbf{r}_b^3w\rb\an\rb e^{i\mbf{r}_b(\mbf{k}_2-\mbf{k}_i)}
    \int d\mbf{r}_c^3w\rc\an\rc e^{i\mbf{r}_c(\mbf{k}_i+\mbf{k}_j-\mbf{k}_1-\mbf{k}_2)}
    \notag\\
    &+
    \int\dfrac{d\mbf{k}^3}{(2\pi)^3}P(\mbf{k})
    \int d\mbf{r}_a^3 w\ra \an\ra e^{i\mbf{r}_a\left(\mbf{k}_1-\mbf{k}\right)}
    \int d\mbf{r}_b^3 w\rb^2\an\rb
    e^{i\mbf{r}_b\left(\mbf{k}-\mbf{k}_1\right)}
    \notag\\
    &+
    \int\dfrac{d\mbf{k}^3}{(2\pi)^3}P(\mbf{k})
    \int d\mbf{r}_b^3 w\rb \an\rb e^{i\mbf{r}_b\left(\mbf{k}_2-\mbf{k}\right)}
    \int d\mbf{r}_c^3 w\rc^2\an\rc
    e^{i\mbf{r}_c\left(\mbf{k}-\mbf{k}_2\right)}
    \notag\\
    &+
    \int\dfrac{d\mbf{k}^3}{(2\pi)^3}P(\mbf{k})
    \int d\mbf{r}_c^3 w\rc \an\rc e^{i\mbf{r}_c\left[-\left(\mbf{k}_2+\mbf{k}_1\right)-\mbf{k}\right]}
    \int d\mbf{r}_a^3 w\ra^2\an\ra
    e^{i\mbf{r}_a\left[\mbf{k}+\left(\mbf{k}_2+\mbf{k}_1\right)\right]}
    \notag\\
    &+
    (1-\alpha^2)\,\mathcal{I}_{31}\Bigg\}
    \notag \\
    &=
    \mathcal{I}^{-1}_{33}\times\Bigg\{
    \int\dfrac{d\mbf{k}_i^3d\mbf{k}_j^3}{(2\pi)^6}
    B(\mbf{k}_i,\mbf{k}_j)\,
    \mathcal{W}^{11}_{\mbf{k}_1-\mbf{k}_j}
    \mathcal{W}^{11}_{\mbf{k}_2-\mbf{k}_i}
    \mathcal{W}^{11}_{\mbf{k}_i+\mbf{k}_j-\mbf{k}_1-\mbf{k}_2}
    \notag\\
    &+
    \int\dfrac{d\mbf{k}^3}{(2\pi)^3}P(\mbf{k})\,
    \left[
    \mathcal{W}^{11}_{\mbf{k}_1-\mbf{k}} 
    \mathcal{W}^{21}_{\mbf{k}-\mbf{k}_1}
    +
    \mathcal{W}^{11}_{\mbf{k}_2-\mbf{k}} 
    \mathcal{W}^{21}_{\mbf{k}-\mbf{k}_2}
    +
    \mathcal{W}^{11}_{-\left(\mbf{k}_2+\mbf{k}_1\right)-\mbf{k}}
    \mathcal{W}^{21}_{\mbf{k}+\left(\mbf{k}_2+\mbf{k}_1\right)}
    \right]
    \notag\\
    &+
    (1-\alpha^2)\,\mathcal{I}_{31}\Bigg\}
    \,,
    \notag \\
\end{align}

\noindent where the notation for the window function was changed as  $\mathcal{W}(\mbf{k})\longrightarrow\mathcal{W}_\mbf{k}$ for a shorthand. In order to subtract the shot-noise component of the measured 3pt correlator in Fourier space using the measured power spectrum (Equation \ref{eq:pk_befassumption}), similarly to what done in Ref. \cite{Gil-Marin:2014sta} we  make the following approximation:

\begin{align}
    \int\dfrac{d\mbf{k}^3}{(2\pi)^3}P(\mbf{k})\,
    \mathcal{W}^{11}_{\mbf{k}_1-\mbf{k}} 
    \mathcal{W}^{21}_{\mbf{k}-\mbf{k}_1}
    \quad \sim \quad
    \dfrac{\mathcal{I}_{32}}{\mathcal{I}_{22}}\int\dfrac{d\mbf{k}^3}{(2\pi)^3}P(\mbf{k})\,
    \mathcal{W}^{11}_{\mbf{k}_1-\mbf{k}} 
    \mathcal{W}^{11}_{\mbf{k}-\mbf{k}_1} \,,
\end{align}

\noindent which is reasonable when not using the FKP weights \cite{Feldman:1993ky} since they usually differs significantly  from unity. With this assumption, it is then possible to write the bispectrum monopole and quadrupole estimators as

\begin{align}
\label{eq:bispectrum_estimator_approx}
\langle F_0(\mbf{k}_1)F_0(\mbf{k}_2)F_0(-\sum_{i=1}^2\mbf{k}_i)\rangle 
&=
\hat{B}^{(0)}(k_1,k_2,k_3)
+ \dfrac{\mathcal{I}_{32}}{\mathcal{I}_{33}}\left[\hat{P}^{(0)}(k_1) +2\,\mrm{cyc.}\right]
+
(1-\alpha^2)\dfrac{\mathcal{I}_{31}}{\mathcal{I}_{33}}
\,,
\notag \\
\langle F_2(\mbf{k}_1)F_0(\mbf{k}_2)F_0(-\sum_{i=1}^2\mbf{k}_i)\rangle 
&=
\hat{B}^{(200)}(k_1,k_2,k_3)
+ \dfrac{\mathcal{I}_{32}}{\mathcal{I}_{33}}\hat{P}^{(2)}(k_1)
\,,
\end{align}

\noindent where the isotropic terms from the bispectrum shot noise's component vanish by definition in the quadrupole's expression. The performance of these estimators in properly accounting for the shot-noise is tested in Figure \ref{fig:PB_shot_noise}.

\subsection{i-Trispectrum}

For the four-point correlation function, in the case of an infinitesimal volume where the number of objects inside a cell can only be $n=0,1$, there are four possibilities:

\paragraph{\texorpdfstring{$\mbf{a\neq b\neq c \neq d}$}{TEXT}:}

\begin{align}
    \langle n_a n_b n_c n_d\rangle=
        \bar{n}\ra\bar{n}\rb\bar{n}\rc\bar{n}\rd \delta V^4
        \Big[1&+\xi(\mbf{r}_{ab}) + 5\quad\mrm{cyc.} 
        + \zeta(\mbf{r}_{ab},\mbf{r}_{bc},\mbf{r}_{ac}) + 3\quad\mrm{cyc.} 
         \notag \\
         &+\xi(\mbf{r}_{ab})\xi(\mbf{r}_{cd}) + 2\quad\mrm{cyc.} 
         + \eta(\mbf{r}_{ab},\mbf{r}_{bc},\mbf{r}_{cd},\mbf{r}_{da})
         \Big]\,.
\end{align}

\paragraph{\texorpdfstring{$\mbf{a = b\neq c \neq d}$}{TEXT}:}
there are six permutations for this term 

\begin{align}
    \langle n_a n_a n_c n_d\rangle= \langle n_a n_c n_d\rangle = 
        \bar{n}\ra\bar{n}\rc\bar{n}\rd \delta V^3
        \left[1+\xi(\mbf{r}_{ac}) + \xi(\mbf{r}_{ad}) + \xi(\mbf{r}_{cd}) + \zeta(\mbf{r}_{ac},\mbf{r}_{ad},\mbf{r}_{cd})\right]\quad.
\end{align}

\paragraph{\texorpdfstring{$\mbf{a = b = c \neq d}$}{TEXT}:}
there are four permutations for this term

\begin{align}
    \langle n_a n_a n_a n_d\rangle= \langle n_a n_d\rangle = 
        \bar{n}\ra\bar{n}\rd \delta V^2
        \left[1+\xi(\mbf{r}_{ad})\right]\quad.
\end{align}

\paragraph{\texorpdfstring{$\mbf{a = b = c = d}$}{TEXT}:}
there is one permutation for this term

\begin{align}
    \langle n_a n_a n_a n_a\rangle= \langle n_a\rangle = 
        \bar{n}\ra \delta V\quad.
\end{align}

\paragraph{Correlators:}
\noindent proceeding as done in Equation \ref{eq:3pt_terms_exp} for the bispectrum, in this case we have:

\begin{align}
\label{eq:4pt_terms_exp}
    \langle n\gal\ra n\gal\rb n\gal\rc n\gal\rd\rangle &= 
    \an\ra\an\rb\an\rc\an\rd
    \Big[1+ \xi\rab + \xi\rac +\xi\rad + \xi\rbc + \xi\rbd + \xi\rcd 
    \notag \\
    &+
    \zeta\rabc + \zeta\rabd + \zeta\racd + \zeta\rbcd +
    \xi\rab\xi\rcd +\xi\rac\xi\rbd +\xi\rad\xi\rbc
    +\eta\rabcd\Big]
    \notag \\
    &+
    \an\ra\an\rc\an\rd\left[1+ \xi\rac+\xi\rad+\xi\rcd + \zeta\racd \right]\delta^D\rab\quad+\quad5\,\mrm{cyc.}
      \notag \\
    &+
    \an\ra\an\rd\left[1+\xi\rad\right]\delta^D\rab\delta^D\rac\quad+\quad3\,\mrm{cyc.}
      \notag \\
    &+
    \an\ra\delta^D\rab\delta^D\rbc\delta^D\rcd
    \notag\\
    \notag\\
        \langle n\gal\ra n\gal\rb n\gal\rc n\rnd\rd\rangle &=
        \alpha^{-1}
    \an\ra\an\rb\an\rc\an\rd
    \Big[1+ \xi\rab + \xi\rac + \xi\rbc + \zeta\rabc\Big]
    \notag \\
    &+
    \alpha^{-1}
    \an\ra\an\rc\an\rd\left[1+ \xi\rac \right]\delta^D\rab\quad+\quad2\,\mrm{cyc.}\quad(\delta^D\rac,\delta^D\rbc)
    \notag \\
    &+
    \alpha^{-1}
    \an\ra\an\rd\delta^D\rab\delta^D\rac
    \notag\\
    \notag\\
    \langle n\gal\ra n\gal\rb n\rnd\rc n\rnd\rd\rangle &=
    \alpha^{-2}
    \an\ra\an\rb\an\rc\an\rd
    \Big[1+ \xi\rab\Big]
    +
    \alpha^{-2}
    \an\ra\an\rc\an\rd\delta^D\rab
    +\alpha^{-1}\an\ra\an\rb\an\rc\delta^D\rcd
    \notag\\
    \notag\\
        \langle n\gal\ra n\rnd\rb n\rnd\rc n\rnd\rd\rangle &=
        \alpha^{-3}
    \an\ra\an\rb\an\rc\an\rd
    +
    \alpha^{-2}
    \an\ra\an\rc\an\rd\delta^D\rbc\quad+\quad2\,\mrm{cyc.}\quad(\delta^D\rbd,\delta^D\rcd)
    \notag \\
    &+
    \alpha^{-1}
    \an\rb\an\rd\delta^D\rbc\delta^D\rbd
    \notag\\
    \notag\\
    \langle n\rnd\ra n\rnd\rb n\rnd\rc n\rnd\rd\rangle &= 
    \alpha^{-4}\an\ra\an\rb\an\rc\an\rd
    +
    \alpha^{-3}\an\ra\an\rc\an\rd\delta^D\rab\quad+\quad5\,\mrm{cyc.}
      \notag \\
    &+
    \alpha^{-2}\an\ra\an\rd\delta^D\rab\delta^D\rac\quad+\quad3\,\mrm{cyc.}
    +
    \alpha^{-1}\an\ra\delta^D\rab\delta^D\rbc\delta^D\rcd
\end{align}

\noindent From the above expression we can derive as done for the bispectrum in Equation \ref{eq:3ptest} the FKP estimator for the trispectrum

\begin{align}
\label{eq:4ptest}
    &\langle F(\mbf{k}_1)F(\mbf{k}_2)F(\mbf{k}_3)F(-\mbf{k}_1-\mbf{k}_2-\mbf{k}_3)\rangle =
    \mathcal{I}^{-1}_{44}\int d\mbf{r}_a^3d\mbf{r}_b^3d\mbf{r}_c^3d\mbf{r}_d^3 w\ra w\rb w\rc w\rd
      \notag \\
    & 
    \times 
    \langle\left[n\ra - \alpha n\rnd\ra\right]\left[n\rb - \alpha n\rnd\rb\right]
    \left[n\rc - \alpha n\rnd\rc\right]\left[n\rd - \alpha n\rnd\rd\right]\rangle 
    e^{i(\mbf{k}_1\mbf{r}_a+\mbf{k}_2\mbf{r}_b+\mbf{k}_3\mbf{r}_c-\mbf{k}_{123}\mbf{r}_d)} \,.
\end{align}

\noindent The following step is to expand the above expression using the terms reported in Equation \ref{eq:4pt_terms_exp}. After simplification the surviving terms are

\begin{align}
\label{eq:4pt_fkp_simplification}
&\langle
\left[n\ra - \alpha n\rnd\ra\right]\left[n\rb - \alpha n\rnd\rb\right]
    \left[n\rc - \alpha n\rnd\rc\right]\left[n\rd - \alpha n\rnd\rd\right]
    \rangle
\notag\\
&=
\na\nb\nc\nd
\Big[\xi\ab\xi\cd+\xi\ac\xi\bd+\xi\ad\xi\bc+\eta\abcd
       \Big]
\notag \\
&+
\na\nb\nc\left[\xi\bc+\zeta\abc\right]\DD\ad
+
\na\nb\nc\left[\xi\ac+\zeta\abc\right]\DD\bd
+
\na\nb\nc\left[\xi\ab+\zeta\abc\right]\DD\cd
\notag \\
&+
\na\nb\nd\left[\xi\bd+\zeta\abd\right]\DD\ac
+
\na\nb\nd\left[\xi\ad+\zeta\abd\right]\DD\bc
+
\na\nc\nd\left[\xi\cd+\zeta\acd\right]\DD\ab
\notag \\
&+
\na\nb\xi\ab\DD\ac\DD\ad + 
\na\nc\xi\ac\DD\ab\DD\ad +
\na\nd\xi\ad\DD\ab\DD\ac +
\na\nb\xi\ab\DD\bc\DD\bd
\notag \\
&+
(1+\alpha^3)\na\DD\ab\DD\ac\DD\ad\,.
\end{align}

\noindent Before proceeding with the simplification of Equation \ref{eq:4pt_fkp_simplification} let's recall the relation between 4pt correlation function $\eta$ and the trispectrum

\begin{align}
    \label{eq:4pt_tk_ft}
    \eta\rabcd=\eta(\mbf{r}_a-\mbf{r}_c,\mbf{r}_b-\mbf{r}_c,\mbf{r}_c-\mbf{r}_d)=
    \int\dfrac{d\mbf{k}_i^3d\mbf{k}_j^3d\mbf{k}_l^3}{(2\pi)^9}
    T(\mbf{k}_i,\mbf{k}_j,\mbf{k}_l)
    e^{-i\left[\mbf{k}_i(\mbf{r}_a-\mbf{r}_c)+\mbf{k}_j(\mbf{r}_b-\mbf{r}_c)+\mbf{k}_l(\mbf{r}_c-\mbf{r}_d)\right]}\,.
\end{align}

\noindent Then we have

\begin{align}
&=
\mathcal{I}^{-1}_{44}\Bigg\{
\int d\mbf{r}_a^3d\mbf{r}_b^3d\mbf{r}_c^3d\mbf{r}_d^3 w\ra w\rb w\rc w\rd
 \na\nb\nc\nd
\eta\abcd
e^{i(\mbf{k}_1\mbf{r}_a+\mbf{k}_2\mbf{r}_b+\mbf{k}_3\mbf{r}_c-\mbf{k}_{123}\mbf{r}_d)}  
\notag \\
&+
\int d\mbf{r}_a^3d\mbf{r}_b^3d\mbf{r}_c^3d\mbf{r}_d^3 w\ra w\rb w\rc w\rd
 \na\nb\nc\nd
\xi\ab\xi\cd
e^{i(\mbf{k}_1\mbf{r}_a+\mbf{k}_2\mbf{r}_b+\mbf{k}_3\mbf{r}_c-\mbf{k}_{123}\mbf{r}_d)}  
\quad+\quad3\;\mr{perms.}
\notag \\
&+
\int d\mbf{r}_a^3d\mbf{r}_b^3d\mbf{r}_c^3w\ra^2w\rb w\rc
\,\na\nb\nc\xi\bc
\times
e^{i(\mbf{k}_1\mbf{r}_a+\mbf{k}_2\mbf{r}_b+\mbf{k}_3\mbf{r}_c-\mbf{k}_{123}\mbf{r}_a)} \,
\quad+\quad5\;\mr{perms.}
\notag \\
&+
\int d\mbf{r}_a^3d\mbf{r}_b^3d\mbf{r}_c^3w\ra^2w\rb w\rc
\,\na\nb\nc\zeta\abc
\times
e^{i(\mbf{k}_1\mbf{r}_a+\mbf{k}_2\mbf{r}_b+\mbf{k}_3\mbf{r}_c-\mbf{k}_{123}\mbf{r}_a)} \,
\quad+\quad5\;\mr{perms.}
\notag \\
&+
\int d\mbf{r}_a^3d\mbf{r}_b^3 w\ra^3w\rb
\na\nb\xi\ab
\times
e^{i(\mbf{k}_1\mbf{r}_a+\mbf{k}_2\mbf{r}_b+\mbf{k}_3\mbf{r}_a-\mbf{k}_{123}\mbf{r}_a)} \,
\quad+\quad3\;\mr{perms.}
\notag \\
&+
\int d\mbf{r}_a^3 w\ra^4
(1+\alpha^3)\na
\times
e^{i(\mbf{k}_1\mbf{r}_a+\mbf{k}_2\mbf{r}_a+\mbf{k}_3\mbf{r}_a-\mbf{k}_{123}\mbf{r}_a)} \,
\,
\Bigg\}
\,.
\end{align}

\noindent Recalling the window and renormalisation factor definitions given in Equation \ref{eq:renormalising_fac} and expanding the $n$-point correlation functions as inverse Fourier transforms of power spectrum, bispectrum and trispectrum, we have that

\begin{align}
\label{eq:tk_fkp_integrals}
    &\langle F(\mbf{k}_1)F(\mbf{k}_2)F(\mbf{k}_3)F(-\mbf{k}_1-\mbf{k}_2-\mbf{k}_3)\rangle 
\notag \\
&=
\mathcal{I}^{-1}_{44}\Bigg\{
\int\dfrac{d\mbf{k}_i^3d\mbf{k}_j^3d\mbf{k}_l^3}{(2\pi)^9}
    T(\mbf{k}_i,\mbf{k}_j,\mbf{k}_l)
\times
\mathcal{W}^{11}_{\mbf{k}_1-\mbf{k}_i}
\mathcal{W}^{11}_{\mbf{k}_2-\mbf{k}_j}
\mathcal{W}^{11}_{\mbf{k}_3-\mbf{k}_l}
\mathcal{W}^{11}_{\mbf{k}_{ijl}-\mbf{k}_{123}}
\notag \\
\notag \\
&+
\int \dfrac{d \mbf{k}^3_i}{(2\pi)^3}\int \dfrac{d \mbf{k}^3_j}{(2\pi)^3}P(k_i)P(k_j)
\times
\mathcal{W}^{11}_{\mbf{k}_1-\mbf{k}_i}
\mathcal{W}^{11}_{\mbf{k}_2+\mbf{k}_i}
\mathcal{W}^{11}_{\mbf{k}_3-\mbf{k}_j}
\mathcal{W}^{11}_{\mbf{k}_j-\mbf{k}_{123}}
\quad+\quad3\;\mr{perms.}
\notag \\
\notag \\
&+
\int\dfrac{d \mbf{k}^3}{(2\pi)^3}  P(k)
\mathcal{W}^{21}_{-\mbf{k}_2-\mbf{k}_3}
\mathcal{W}^{11}_{\mbf{k}_2-\mbf{k}}
\mathcal{W}^{11}_{\mbf{k}_3+\mbf{k}}
\quad+\quad5\;\mr{perms.}
\notag \\
\notag \\
&+
\int\dfrac{d\mbf{k}_i^3d\mbf{k}_j^3}{(2\pi)^6}   
B(\mbf{k}_i,\mbf{k}_j)
\mathcal{W}^{21}_{-\mbf{k}_{23}-\mbf{k}_i}
\mathcal{W}^{11}_{\mbf{k}_2-\mbf{k}_j}
\mathcal{W}^{11}_{\mbf{k}_3+\mbf{k}_{ij}}
\quad+\quad5\;\mr{perms.}
\notag \\
\notag \\
&+
\int\dfrac{d \mbf{k}^3}{(2\pi)^3} P(k)
\mathcal{W}^{31}_{-\mbf{k}_2-\mbf{k}}
\mathcal{W}^{11}_{\mbf{k}_2+\mbf{k}}
\quad+\quad3\;\mr{perms.}
\notag \\
\notag \\
&+
(1+\alpha^3)\mathcal{I}_{41}
\Bigg\}\,
\,.
\end{align}

\noindent As done for the bispectrum in Equation \ref{eq:bispectrum_estimator_approx}, for the i-trispectrum in order to subtract the shot-noise using measured quantities we need to assume in addition that:

\begin{align}
 \int\dfrac{d \mbf{k}^3}{(2\pi)^3}  P(k)
\mathcal{W}^{21}_{-\mbf{k}_2-\mbf{k}_3}
\mathcal{W}^{11}_{\mbf{k}_2-\mbf{k}}
\mathcal{W}^{11}_{\mbf{k}_3+\mbf{k}}
\;&\sim\; 
\dfrac{\mathcal{I}_{42}}{\mathcal{I}_{22}}
\int\dfrac{d \mbf{k}^3}{(2\pi)^3}  P(k)
\mathcal{W}^{11}_{\mbf{k}-\mbf{k}_2-\mbf{k}_3}
\mathcal{W}^{11}_{\mbf{k}-\mbf{k}_2-\mbf{k}_3}
\notag \\
\int\dfrac{d\mbf{k}_i^3d\mbf{k}_j^3}{(2\pi)^6}   
B(\mbf{k}_i,\mbf{k}_j)
\mathcal{W}^{21}_{-\mbf{k}_{23}-\mbf{k}_i}
\mathcal{W}^{11}_{\mbf{k}_2-\mbf{k}_j}
\mathcal{W}^{11}_{\mbf{k}_3+\mbf{k}_{ij}}
\;&\sim\;
\dfrac{\mathcal{I}_{43}}{\mathcal{I}_{33}}
\int\dfrac{d\mbf{k}_i^3d\mbf{k}_j^3}{(2\pi)^6}   
B(\mbf{k}_i,\mbf{k}_j)
\mathcal{W}^{11}_{-\mbf{k}_{23}-\mbf{k}_i}
\mathcal{W}^{11}_{\mbf{k}_2-\mbf{k}_j}
\mathcal{W}^{11}_{\mbf{k}_3+\mbf{k}_{ij}}
\notag \\
\int\dfrac{d \mbf{k}^3}{(2\pi)^3} P(k)
\mathcal{W}^{31}_{-\mbf{k}_2-\mbf{k}}
\mathcal{W}^{11}_{\mbf{k}_2+\mbf{k}}
\;&\sim\;
\dfrac{\mathcal{I}_{42}}{\mathcal{I}_{22}}
\int\dfrac{d \mbf{k}^3}{(2\pi)^3} P(k)
\mathcal{W}^{11}_{-\mbf{k}_2-\mbf{k}}
\mathcal{W}^{11}_{\mbf{k}_2+\mbf{k}}
\,,
\notag \\
\end{align}

\noindent where in particular the first approximation is driven by what theoretical derivations using generating functionals in \cite{Verde:2001pf} and our previous analyses in the case of simulations with periodic boundary conditions \cite{Gualdi:2020eag,Gualdi:2021yvq}.  Then our estimators for monopole and quadrupoles of the i-trispectrum become

\begin{align}
\langle F_0(\mbf{k}_1)F_0(\mbf{k}_2)F_0(\mbf{k}_3)F_0(-\sum_{i=1}^3\mbf{k}_i)\rangle 
&=
\hat{\mathcal{T}}^{(0)}(k_1,k_2,k_3,k_4)
\notag \\
+
\dfrac{1}{N_{D_1}N_{D_2}}
\sum^{N_{D_1}}_i
\sum^{N_{D_2}}_j
\Bigg\{
\dfrac{\mathcal{I}_{43}}{\mathcal{I}_{44}}
\Big[
\hat{B}^{(0)}(k_1,k_2,D_{1,i}) &+ 
\hat{B}^{(0)}(k_1,k_3,|\mbf{k}_2+\mbf{k}_4|) + 
\hat{B}^{(0)}(k_1,k_4,D_{2,j}) 
\notag \\
+
\hat{B}^{(0)}(k_2,k_3,D_{2,j}) &+ 
\hat{B}^{(0)}(k_2,k_4,|\mbf{k}_1+\mbf{k}_3|) + 
\hat{B}^{(0)}(k_3,k_4,D_{1,i})
\Big]
\notag \\
+\dfrac{\mathcal{I}_{42}}{\mathcal{I}_{44}}
\Big[
\hat{P}^{(0)}(k_1)+\hat{P}^{(0)}(k_2)&+
\hat{P}^{(0)}(k_3)+
\hat{P}^{(0)}(k_4)
\notag \\ 
+\hat{P}^{(0)}(D_{1,i})+\hat{P}^{(0)}(D_{2,j})&+
\hat{P}^{(0)}(|\mbf{k}_1+\mbf{k}_3|)
\Big]
\Bigg\}
+
(1+\alpha^3)\dfrac{\mathcal{I}_{41}}{\mathcal{I}_{44}}
\,,
\notag \\
\langle F_2(\mbf{k}_1)F_0(\mbf{k}_2)F_0(\mbf{k}_3)F_0(-\sum_{i=1}^3\mbf{k}_i)\rangle 
&=
\hat{\mathcal{T}}^{(2000)}(k_1,k_2,k_3,k_4)
\notag \\
+
\dfrac{1}{N_{D_1}N_{D_2}}
\sum^{N_{D_1}}_i
\sum^{N_{D_2}}_j
\Bigg\{
\dfrac{\mathcal{I}_{43}}{\mathcal{I}_{44}}
\Big[
\hat{B}^{(200)}(k_1,k_2,D_{1,i}) &+ 
\hat{B}^{(200)}(k_1,k_3,|\mbf{k}_2+\mbf{k}_4|) + 
\hat{B}^{(200)}(k_1,k_4,D_{2,j}) 
\notag \\
+
\hat{B}^{(002)}(k_2,k_3,D_{2,j}) &+ 
\hat{B}^{(002)}(k_2,k_4,|\mbf{k}_1+\mbf{k}_3|) + 
\hat{B}^{(002)}(k_3,k_4,D_{1,i})
\Big]
\notag \\
+\dfrac{\mathcal{I}_{42}}{\mathcal{I}_{44}}
\Big[
\hat{P}^{(2)}(k_1) +\hat{P}^{(2)}(D_{1,i})&+
\hat{P}^{(2)}(D_{2,j})+\hat{P}^{(2)}(|\mbf{k}_1+\mbf{k}_3|)
\Big]
\Bigg\}\,,
\end{align}

\noindent where it becomes essential the summation over all the possible values of the diagonals $D_1$ and $D_1$ when passing from the integrals to the discrete sums. Notice that in the shot-noise term for $\mathcal{T}^{(0)}$ the permutations relative to the term $P^{(0)}\left(\mbf{k}_i+\mbf{k}_j\right)$ are only three, differently from Equation \ref{eq:tk_fkp_integrals} where there were six. This is because of the symmetries of the skew-quadrilaterals: there are only three different possible sums of two $k$-vectors out of the set of four. As for the power spectrum and bispectrum, in the quadrupoles estimators all the isotropic terms vanish.
\section{Results for power spectrum and bispectrum}
\label{sec:p_b_results}

\begin{figure}[tbp]
\centering 
\includegraphics[width=1.\textwidth]
{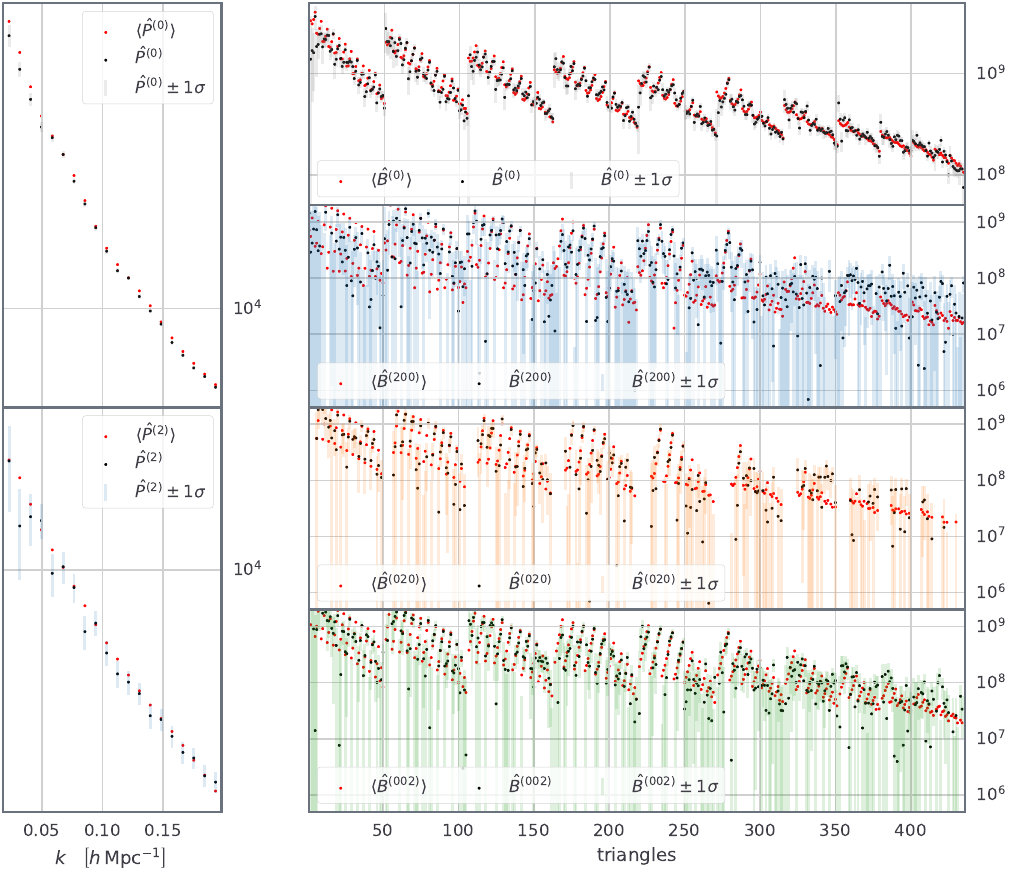}
\caption{\label{fig:PB_det}
Measurements of the power spectrum and bispectrum, monopole and quadrupoles, from BOSS DR12 CMASS NGC data. The errorbars are obtained from the covariance estimated using 2048 realisations of the Patchy Mocks and are centered around the black dots which are the statistics measurements from BOSS data. For comparison the red dots show the average measurements of the same quantities from the Patchy Mocks.
}
\end{figure}

\begin{figure}[tbp]
\centering 
\includegraphics[width=1.\textwidth]
{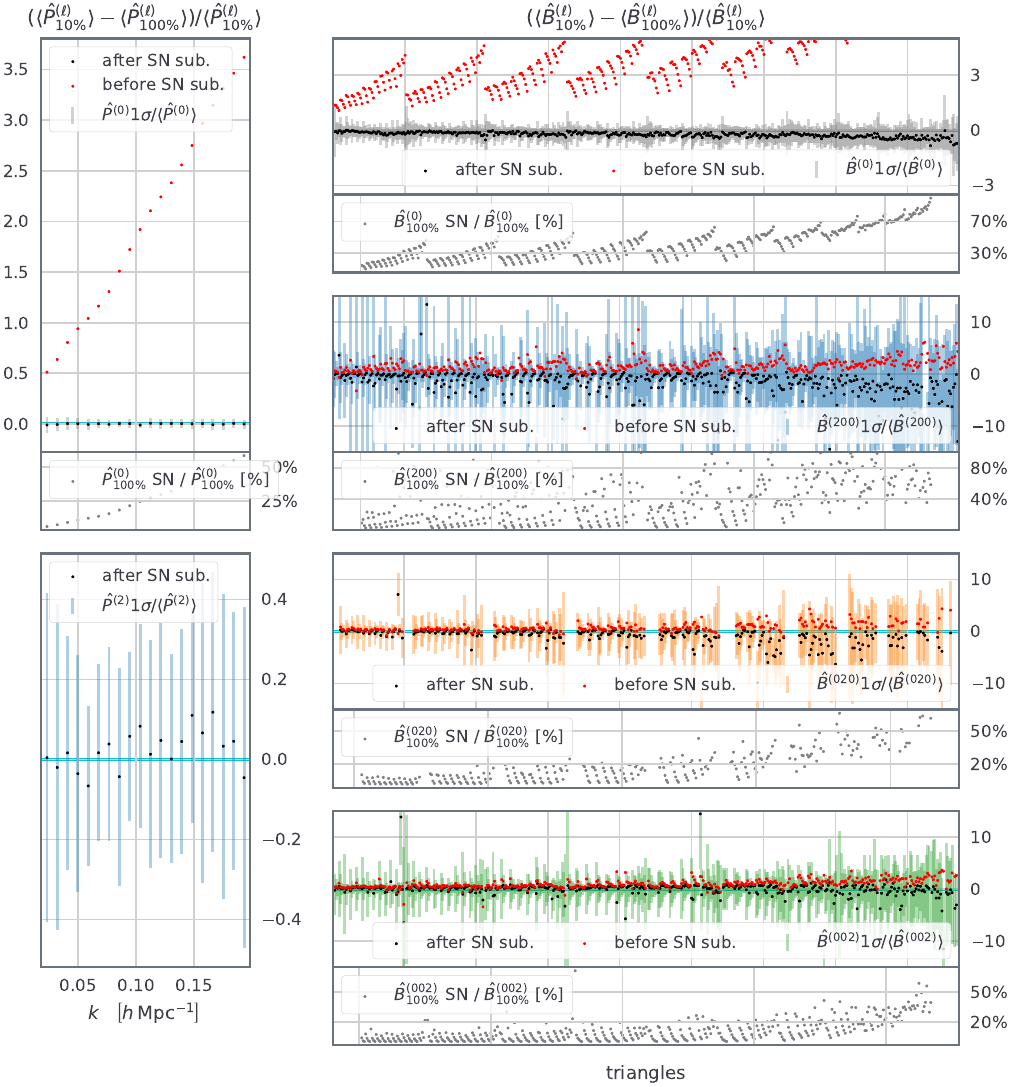}
\caption{\label{fig:PB_shot_noise}
Analogous of Figure \ref{fig:T_shot_noise_test} but for power spectrum and bispectrum multipoles. In the case of $P^{(2)}$ the shot-noise correction is equal to zero by definition, therefore in the relative plot just the ratio between full-density and undersampled catalogues measurements is shown. 
}
\end{figure}

\begin{figure}[tbp]
\centering 
\includegraphics[width=1.\textwidth]
{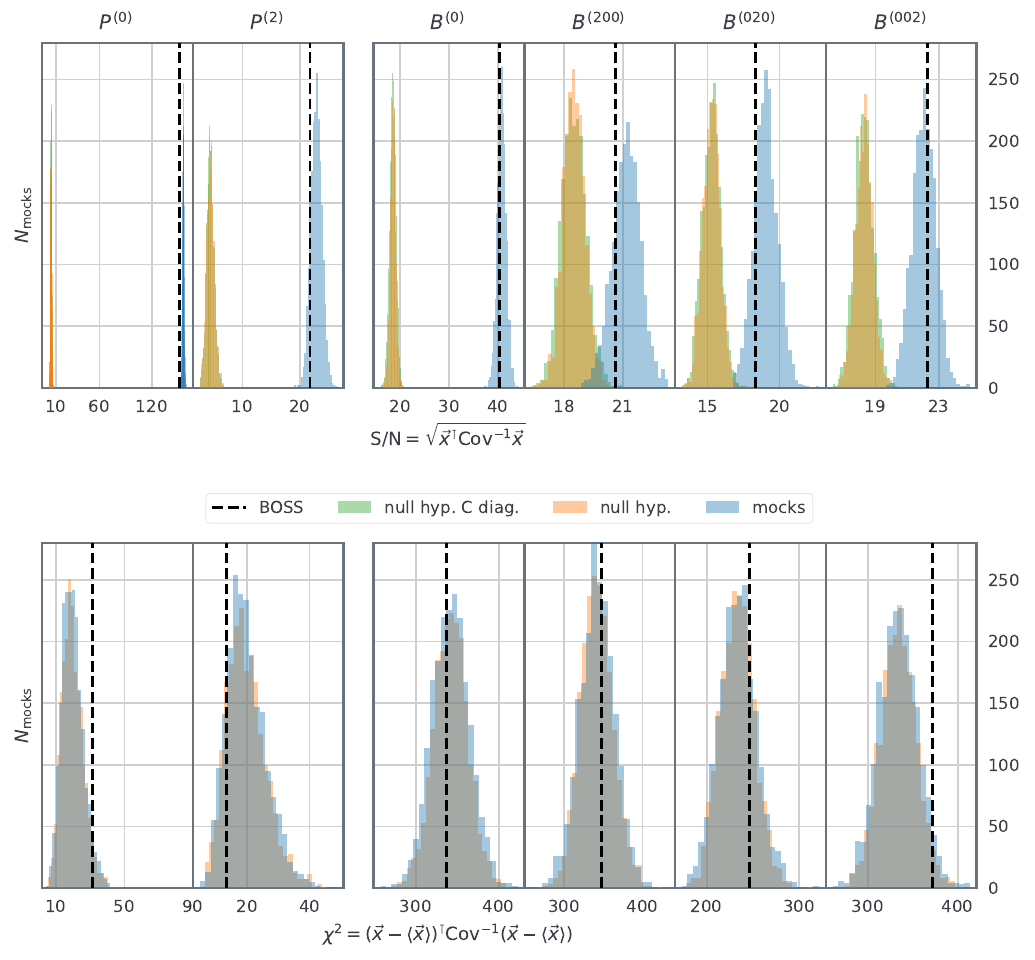}
\caption{\label{fig:PB_ston_corr_chi2}
Analogous of Figure \ref{fig:T_ston_corr_chi2} but for power spectrum and bispectrum multipoles.
In the power spectrum case the relative error (in particular for the monopole) is much smaller than for bispectrum and i-trispectrum. This is why the discrepancy between measurements from data and galaxy catalogues becomes more significant in terms of $\chi^2$ for $P^{(0,2)}$ as it is evident from the figure's second row.}
\end{figure}

In this Appendix the  results presented in the paper for the i-trispectrum are instead reported for both power spectrum and bispectrum, mainly in the form of plots. In Figure \ref{fig:PB_det} are shown the measurements from BOSS CMASS NGC data for power spectrum and bispectrum multipoles. In Figure \ref{fig:PB_shot_noise} we check the shot-noise subtraction for the power spectrum and bispectrum estimator by comparing measurements from catalogues with different densities as described in Section \ref{sec:shot_noise_test}.
Finally in Figure \ref{fig:PB_ston_corr_chi2} we show the detection for power spectrum and bispectrum in terms of signal-to-noise ratio and $\chi^2$ test with respect to the mocks average measurements.
Table \ref{tab:ston_pb} shows that while for power spectrum multipoles and bispectrum monopole the detection's significance is higher than for the i-trispectrum monopole, the distance between physical signal and null hypothesis is similar for what concerns bispectrum and i-trispectrum quadrupoles.

\section{Quadrilaterals selection}
\label{sec:remove_quads}

\begin{figure}[tbp]
\centering 
\includegraphics[width=1.\textwidth]
{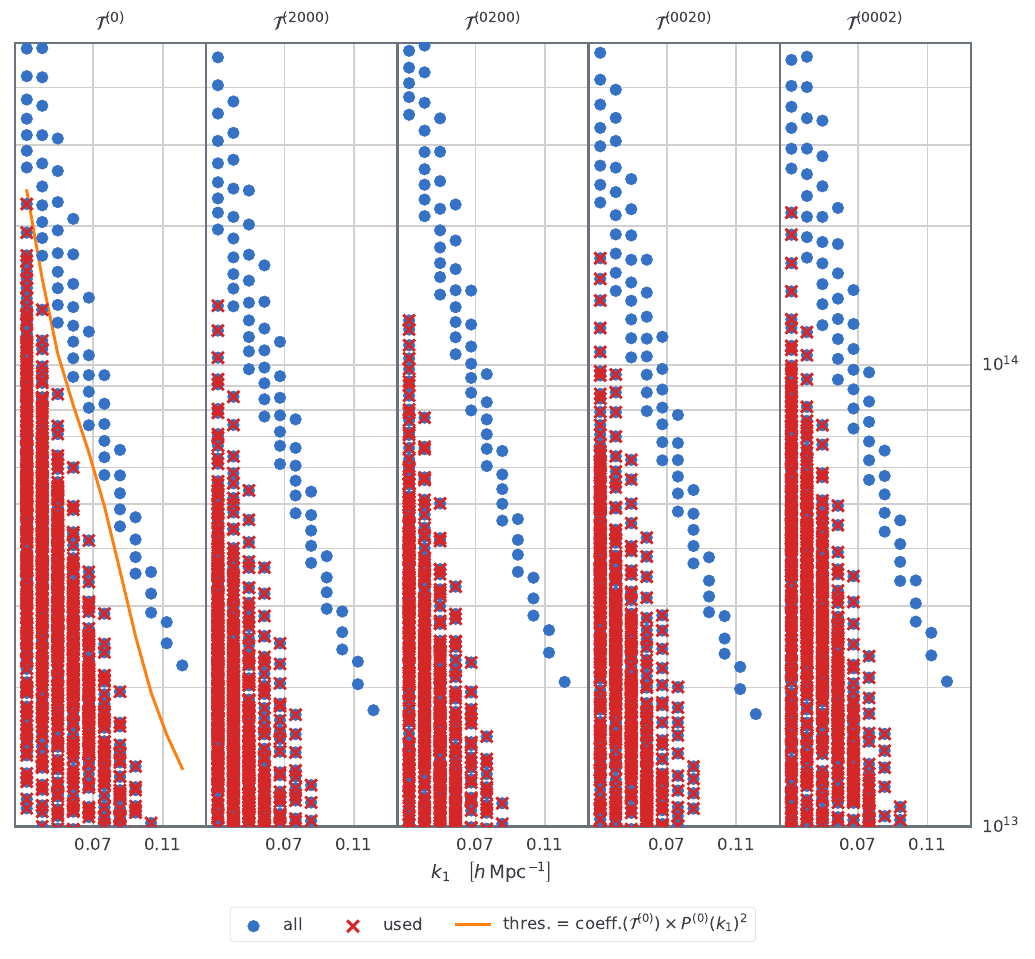}
\caption{\label{fig:sel_quadrilaterals}
I-trispectrum monopole and quadrupoles signal as a function of the shortest side $k_1$. Blue circles show the signal for all the measured configurations, the red "x's" indicate the remaining configurations after applying the selection criteria. The latter is given by the orange line which uses a simple prescription (Equation \ref{eq:rem_quads}) to remove the configurations whose signal is dominated by the convolution with the survey mask, creating pseudo-unconnected four-point correlator terms in Fourier space \cite{Gualdi:2021yvq}. The empirical value adopted for the coefficient is $\mrm{coeff.}(\mathcal{T}^{(0)})=9\times10^4$.
}
\end{figure}

\begin{figure}[tbp]
\centering 
\includegraphics[width=1.\textwidth]
{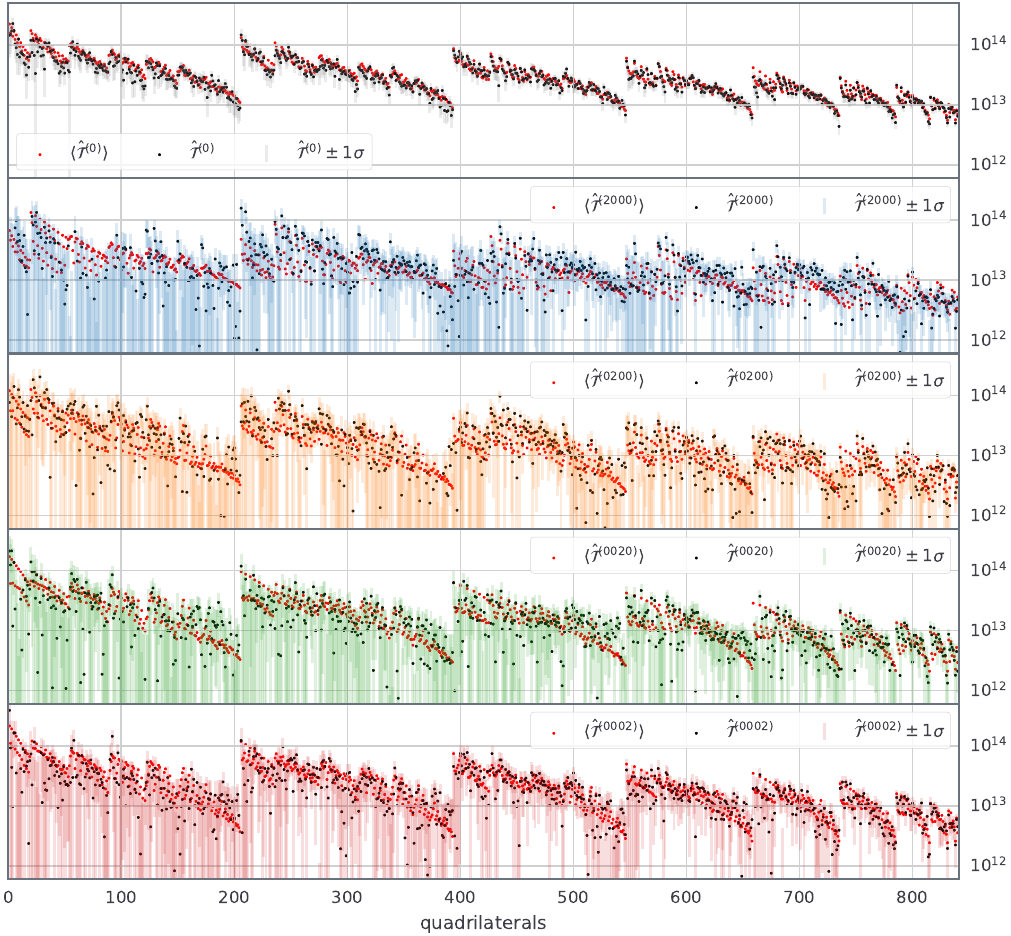}
\caption{\label{fig:T_det_sel_quads}
Equivalent of Figure \ref{fig:T_det_all_quads} after the removal of the quadrilaterals whose i-trispectrum signal is dominated by the convolution with the survey window function (Figure \ref{fig:sel_quadrilaterals}).
}
\end{figure}

When measuring statistics in Fourier space from surveys, the physical signal is convolved with the surveys window function which encode the non-regularity of the observed volume and hence the lack of periodicity in its boundary conditions. When performing parameters constraints analyses, accounting for this effect at the level of the power spectrum has been done either by also convolving the theoretical model with the window function \cite{Wilson:2015lup} or by deconvolving the measurements from it \cite{Beutler:2021eqq}, while for the bispectrum a first attempt to account for the mask was presented in \cite{Gil-Marin:2016wya}. For the i-trispectrum's detection we remove configurations that are clearly dominated by the convolution with the window function, which can be seen by eye in Figure \ref{fig:T_det_all_quads} by comparison with our previous work \cite{Gualdi:2020eag,Gualdi:2021yvq}. This can be done using a simple empirical prescription as shown in Figure \ref{fig:sel_quadrilaterals} which consists in removing all the configurations whose i-trispectrum monopole signal is larger than a pseudo-unconnected component's signal for the smallest $k$-mode (i.e. largest power spectrum) with an arbitrary coefficient:

\begin{eqnarray}
\label{eq:rem_quads}
\mrm{thres.} = \mrm{coeff.}(\mathcal{T}^{(0)}) \times P^{(0)}(k_1)^2\,,
\end{eqnarray}

\noindent where in our case we use $\mrm{coeff.}(\mathcal{T}^{(0)})=9\times10^4$. In this work the selection criteria derived for the mocks is also applied to the data. The i-trispectrum multipoles data-vector measured from data after quadrilateral selection is shown in Figure \ref{fig:T_det_sel_quads}. We anticipate however, that modeling the effect of the window function on the signal  will be the next challenge to be overcome before the trispectrum signal can be interpreted in light of theory and used to constrain (cosmological) parameters.

\renewcommand{\arraystretch}{2.}
\begin{table}[tbp]
\centering
\begin{tabular}{c|c|c|c|c|c|c|}
\cline{2-7}
 & \multicolumn{6}{c|}{$S/N$ and $(\Delta \sigma)$-distance } \\
\cline{2-7}
& 2.28 $\%$ & 15.87 $\%$ & 50 $\%$ & 84.14 $\%$ & 97.73 $\%$ & DATA\\
\hline
$\mathcal{P}^{(0)}$   & \textbf{153 (212.5)} & 154 (214)            & 155 (215)            & 156.4 (217)         & 157.5 (218) & \textbf{152 (211)} \\
$\mathcal{P}^{(2)}$   & \textbf{20.9 (24.3)} & \textbf{21.9 (25.7)} & 23.9 (27.2)          & 24.8 (28.7)         & 24.7 (30.1) & \textbf{21.8 (25.6)} \\
\hline
$\mathcal{B}^{(0)}$   & 39.0 (37.7)          & \textbf{39.9 (39.4)} & \textbf{40.8 (41.0)} & 41.7 (42.7)         & 42.6 (44.4) & \textbf{40.3 (40.0)} \\
$\mathcal{B}^{(200)}$ & 19.7 (2.1)           & \textbf{20.5 (3.4)}  & \textbf{21.2 (4.8)}  & 21.9 (6.0)          & 22.6 (7.3)  & \textbf{20.6 (3.7)} \\
$\mathcal{B}^{(020)}$ & 17.4 (3.6)           & \textbf{18.2 (5.0)}  & \textbf{19.0 (6.3)}  & 19.9 (7.7)          & 20.6 (9.0)  & \textbf{18.3 (5.1)} \\
$\mathcal{B}^{(002)}$ & 20.6 (4.2)           & 21.3 (5.6)           & \textbf{22.1 (7.0)}  & \textbf{22.8 (8.3)} & 23.5 (9.6)  & \textbf{22.3 (7.3)} \\
\hline
\end{tabular}
\caption{\label{tab:ston_pb}
Analogous of Table \ref{tab:ston_t} for the power spectrum and bispectrum monopole and quadrupoles.
}
\end{table}

\bibliography{references}
\bibliographystyle{utcaps}


\end{document}